\documentclass[sigconf]{acmart}

\AtBeginDocument{%
  \providecommand\BibTeX{{%
    \normalfont B\kern-0.5em{\scshape i\kern-0.25em b}\kern-0.8em\TeX}}}

\usepackage{booktabs,multirow}
\usepackage{subfigure} 
\usepackage{balance}
\usepackage{graphicx}
\usepackage{caption}
\usepackage{algorithm}
\usepackage{algpseudocode}  
\usepackage{multirow}
\usepackage{bm}

\usepackage{orcidlink}
\copyrightyear{2025}
\acmYear{2025}
\setcopyright{cc}
\setcctype{by}
\acmConference[KDD '25]{Proceedings of the 31st ACM SIGKDD Conference on Knowledge Discovery and Data Mining V.2}{August 3--7, 2025}{Toronto, ON, Canada}
\acmBooktitle{Proceedings of the 31st ACM SIGKDD Conference on Knowledge Discovery and Data Mining V.2 (KDD '25), August 3--7, 2025, Toronto, ON, Canada}
\acmDOI{10.1145/3711896.3737087}
\acmISBN{979-8-4007-1454-2/2025/08}

\makeatletter
\gdef\@copyrightpermission{
  \begin{minipage}{0.3\columnwidth}
   \href{https://creativecommons.org/licenses/by/4.0/}{\includegraphics[width=0.90\textwidth]{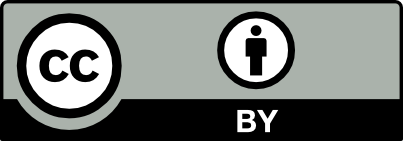}}
  \end{minipage}\hfill
  \begin{minipage}{0.7\columnwidth}
   \href{https://creativecommons.org/licenses/by/4.0/}{This work is licensed under Creative Commons Attribution International 4.0.}
  \end{minipage}
  \vspace{5pt}
}
\makeatother

\begin{document}

\title{Predicting the Dynamics of Complex System via Multiscale Diffusion Autoencoder}

\author{Ruikun Li}
\affiliation{
  \institution{Shenzhen International Graduate School, Tsinghua University}
  \state{Shenzhen}
  \country{China}
}

\author{Jingwen Cheng}
\affiliation{
  \institution{Department of Electronic Engineering\\ BNRist, Tsinghua University}
  \state{Beijing}
  \country{China}
}

\author{Huandong Wang}
\authornote{Huandong Wang is the corresponding author (wanghuandong@tsinghua.edu.cn).}
\affiliation{
  \institution{Department of Electronic Engineering\\ BNRist, Tsinghua University}
  \state{Beijing}
  \country{China}
}

\author{Qingmin Liao}
\affiliation{
  \institution{Shenzhen International Graduate School, Tsinghua University}
  \state{Shenzhen}
  \country{China}
}

\author{Yong Li}
\affiliation{
  \institution{Department of Electronic Engineering\\ BNRist, Tsinghua University}
  \state{Beijing}
  \country{China}
}

\begin{abstract}
  Predicting the dynamics of complex systems is crucial for various scientific and engineering applications. 
  The accuracy of predictions depends on the model's ability to capture the intrinsic dynamics. 
  While existing methods capture key dynamics by encoding a low-dimensional latent space, they overlook the inherent multiscale structure of complex systems, making it difficult to accurately predict complex spatiotemporal evolution.
  Therefore, we propose a \underline{M}ultiscale  \underline{D}iffusion  \underline{P}rediction  \underline{Net}work (MDPNet) that leverages the multiscale structure of complex systems to discover the latent space of intrinsic dynamics.
  First, we encode multiscale features through a multiscale diffusion autoencoder to guide the diffusion model for reliable reconstruction.
  Then, we introduce an attention-based graph neural ordinary differential equation to model the co-evolution across different scales.
  Extensive evaluations on representative systems demonstrate that the proposed method achieves an average prediction error reduction of 53.23\% compared to baselines, while also exhibiting superior robustness and generalization.
\end{abstract}




\maketitle

\newcommand\kddavailabilityurl{https://doi.org/10.5281/zenodo.15574207}

\ifdefempty{\kddavailabilityurl}{}{
\begingroup\small\noindent\raggedright\textbf{KDD Availability Link:}
The source code of this paper has been made publicly available at \url{\kddavailabilityurl}.
\endgroup
}

\section{Introduction}

The dynamics of complex systems emerge from the nonlinear interactions and co-evolution of numerous components, giving rise to intricate spatiotemporal patterns and multiscale structures, as seen in fluid flow~\cite{kochkov2021machine}, bioproteins~\cite{bhatia2021machine}, and brain neurons~\cite{brooks2009charmm}.
Predicting the dynamics of such systems is crucial for real-world applications, including policy-making, resource management, and strategic planning~\cite{preisler2007statistical, brown2015future, bevacqua2023advancing}. 
% However, direct numerical simulations rely heavily on strong prior knowledge of the system, limiting their practical applicability~\cite{lifourier, ruhling2024dyffusion}.
% A promising approach is to uncover the intrinsic dynamics hidden within high-dimensional observational spaces, enabling data-driven modeling that circumvents the dependence on governing equations\cite{vlachas2022multiscale, gao2024generative, wu2024neural, lu2021learning}. 
% The key challenge, however, lies in identifying an appropriate latent space where the inherent properties of the physical system are preserved while ensuring accurate dynamical predictions~\cite{wang2024multi, gao2024intrinsic, thibeault2024low}.
In previous studies, the remarkable complexity of these systems has shown potential for reduction through dimensionality reduction techniques~\cite{gao2024intrinsic, li2025predicting}. 
The latent variables that bridge high-dimensional system states are sufficient to describe the emergent insightful phenomena arising from complex microscopic dynamics~\cite{thibeault2024low}. 
Guided by this idea, a core objective of complex system dynamics prediction is to discover the low-dimensional latent space where the system's intrinsic dynamics reside~\cite{vlachas2022multiscale, kontolati2024learning, wu2024neural,ding2025artificial}.

\begin{figure}[!t]
    \centering
    \includegraphics[width=0.85\linewidth]{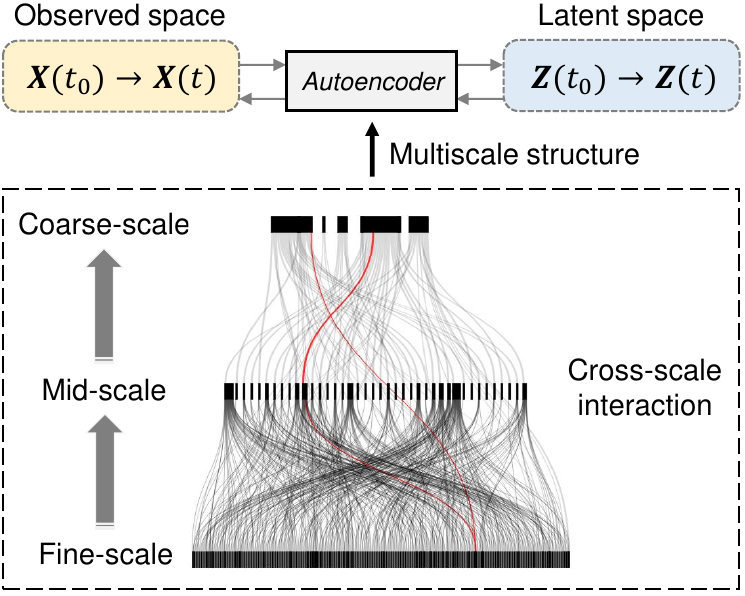}
    \caption{Latent representation based on multiscale structure.}
    \label{fig:multiscale}
\end{figure}

As a long-standing problem, numerous methods have been developed to model the intrinsic dynamics of complex systems, including physics-based methods~\cite{li2004multi} and machine learning methods~\cite{wang2024multi}.
Traditional physics-based methods identify low-dimensional spaces by applying subsampling or coarse-graining to system states at different scales.
\citet{bar2019learning} averages out fast-scale dynamical features to derive coarse-grained dynamics on a low-resolution grid, an idea that can be traced back to the slaving principle of Haken's synergetics~\cite{haken2012advanced}.
\citet{gao2024generative} performs nonparametric subsampling on high-dimensional states to construct a coarse-scale latent space.
Moreover, renormalization group-based methods~\cite{garcia2018multiscale, koch2018mutual, villegas2023laplacian} have been developed, where nodes are aggregated into supernodes, iteratively forming coarse-grained structures of networked systems at different scales.
These physics-based methods effectively leverage the inherent multiscale structure of complex systems. 
However, their rule-based subsampling or coarse-graining inevitably lead to the loss of fine-scale details, limiting predictive accuracy.
By contrast, deep learning methods using autoencoders avoid information loss during encoding by minimizing self-supervised reconstruction errors.
\citet{vlachas2022multiscale} designs an appropriate bottleneck dimension for the autoencoder to faithfully encode the low-dimensional latent space of the dynamics.
Recent studies have explored the discovery of physically meaningful low-dimensional latent representations using physics-informed loss functions~\cite{lusch2018deep, li2023learning} or physics-inspired embedding methods~\cite{wu2024predicting}.
The aforementioned autoencoder-based methods treat neural networks as black boxes, mapping the system into a latent space where global information is entangled.
Despite benefiting from the strong fitting capabilities of deep learning models, these methods do not account for the inherent multiscale structure of complex systems, limiting their predictive power for intricate spatiotemporal patterns.
Taken together, whether an appropriate latent space can be found that preserves the multiscale structure of the system while avoiding information loss remains an open question, as illustrated in Figure~\ref{fig:multiscale}.

Effectively modeling the latent space with multiscale structure presents two key challenges.
First, it is difficult to incorporate information from different scales when mapping between the observational space and the latent space.
Second, effectively capturing cross-scale interactions remains challenging, as information propagates across scales in a highly nonlinear and dynamic manner.
Recent studies have demonstrated that diffusion models excel at capturing complex spatiotemporal distributions~\cite{gao2024generative, ruhling2024dyffusion, li2024learning}. 
The sampling process of diffusion models gradually denoises a standard normal distribution into a data distribution, which naturally aligns with the progressive transition of data structures from coarse to fine scales~\cite{fan2024mg, shen2024multi, jin2024pyramidal}.
This inherent coarse-to-fine transition suggests that diffusion models can serve as a structured generative framework for modeling the multiscale structure of complex systems.
However, a major challenge remains in decoupling multiscale representations and predicting their co-evolution, which standard diffusion models do not inherently address.

To address the challenges mentioned above, we propose a novel deep learning framework, named \underline{M}ultiscale \underline{D}iffusion \underline{P}rediction \underline{Net}work (MDPNet), for predicting intrinsic dynamics in the latent space. 
MDPNet first maps the system state to latent spaces of different scales using a multiscale residual encoder. The coarsening-guided diffusion decoder then reconstructs the original observations by taking these scale-specific latent vectors as conditional inputs. 
Together, they form a novel multiscale diffusion autoencoder, which treats encoded vectors as multiscale guidance conditions for the diffusion process to collaboratively uncover a low-dimensional latent space of multiscale structures. This addresses the first challenge. 
Furthermore, we connect each scale in the latent space and design a graph neural ordinary differential equation based on the graph attention mechanism to automatically aggregate cross-scale information propagation, modeling the co-evolutionary dynamics across scales and overcoming the second challenge.

Our contribution can be summarized as follows:
\begin{itemize}
    \item We introduce a multiscale diffusion autoencoder that combines the multiscale structure of complex systems with deep representation learning, which decouples and preserves multiscale information.
    % \item We propose a coarsening-guided diffusion decoder to enhance the fitting ability of standard diffusion models and collaborate with the encoder to uncover the multiscale latent space of complex systems.
    \item We develop an attention-based graph neural differential equation to automatically model the cross-scale interaction of complex systems, enabling accurate predictions of the co-evolution across scales.
    \item Extensive experiments on four representative systems show that MDPNet outperforms state-of-the-art baselines by an average of 53.23\% in prediction error. 
    % The code is available at: \href{https://github.com/tsinghua-fib-lab/MDPNet}{https://github.com/tsinghua-fib-lab/MDPNet}.
\end{itemize}
\section{Preliminary}

\subsection{Problem Definition}

We study the problem of spatiotemporal prediction given an initial condition. The data is collected as a set of snapshots $\{\bm{x}_t\}^T_{t=1}$, where $\bm{x}_t \in \mathbb{R}^{C \times H \times W}$, $H \times W$ represents the 2-dim spatial grid, and the channel $C$ corresponds to the physical variables defined on the grid. We focus on the conditional probability distribution $\bm{p}(\bm{x}_\tau | \bm{x}_0)$ of the system state at a future time $\tau$, given an initial condition $\bm{x}_0$. 
% To accurately predict the long-term evolution of the system, the model must effectively capture the system's intrinsic dynamics rather than merely extrapolating statistical averages.

\subsection{Diffusion Model}\label{sec:pre_diff}

Diffusion model~\cite{dhariwal2021diffusion, rombach2022high} perturbs the data distribution by adding noise and learn to reverse the process through denoising, demonstrating strong fitting capabilities for data distributions in images, videos, and general sequences~\cite{yang2023diffusion, cao2024survey, croitoru2023diffusion, sheng2025collaborative}. 
We denote the original sample as $\bm{x}_0$ and the sample after $n$ diffusion steps as $\bm{x}_n$.
In a standard diffusion model, the forward diffusion process iteratively adds noise to perturb the data: $\bm{x}_n=\sqrt{\overline{\alpha}_n}\bm{x}_0+\sqrt{1-\overline{\alpha}_n}\bm{\epsilon}$, where $\bm{\epsilon} \sim \mathcal{N}(0,\bm{I})$ and $\{\alpha_n\}$ are noise schedules~\cite{ho2020denoising}.
The reverse process starts from Gaussian noise and progressively denoising to sample the real data point as
\begin{equation}
    p_\theta(\bm{x}_{n-1}|\bm{x}_n):=\mathcal{N}(\bm{x}_{n-1}; \bm{\mu}_\theta(\bm{x}_n,n),\sigma^2_n\bm{I}),
\end{equation}
where $\bm{\mu}_\theta=\frac{1}{\sqrt{\alpha_n}}(\bm{x}_n-\frac{1-\alpha_n}{\sqrt{1-\overline{\alpha}_n}}\epsilon_\theta(\bm{x}_n,n))$ and $\{\sigma_n\}$ are step dependent constants. Noise $\bm{\epsilon}_\theta$ represents the single-step noise estimated by the parameterized neural network (also referred to as the score function \cite{song2019generative,song2020score}), which is typically formalized as a UNet architecture \cite{song2020denoising, karras2022elucidating}. 
Parameters of networks can be optimized using the objective function~\cite{ho2020denoising}
\begin{equation}
    L_n=\mathbb{E}_{n,\bm{\epsilon}_n,\bm{x}_0}||\bm{\epsilon}_n-\bm{\epsilon}_\theta(\sqrt{\overline{\alpha}_n}\bm{x}_0+\sqrt{1-\overline{\alpha}_n}\bm{\epsilon}_n, n)||^2
\end{equation}
to minimize the negative log-likelihood $\mathbb{E}_{\bm{x}_0 \sim \bm{q}(\bm{x}_0)}[-\bm{p}_\theta(\bm{x}_0)]$.
Once trained, the model can unconditionally generate diverse samples by repeatedly sampling Gaussian noise and executing the reverse process. To ensure that the generated content aligns with the prompt, the fields of video generation and time-series modeling commonly incorporate cross-attention modules~\cite{tumanyan2023plug, rombach2022high} into the UNet architecture, allowing external conditional information $\bm{c}$ to guide noise estimation, $\bm{\epsilon}_\theta(\bm{x}_n,n,\bm{c})$.
\section{Methodology}

\begin{figure*}[!t]
    \centering
    \includegraphics[width=0.9\linewidth]{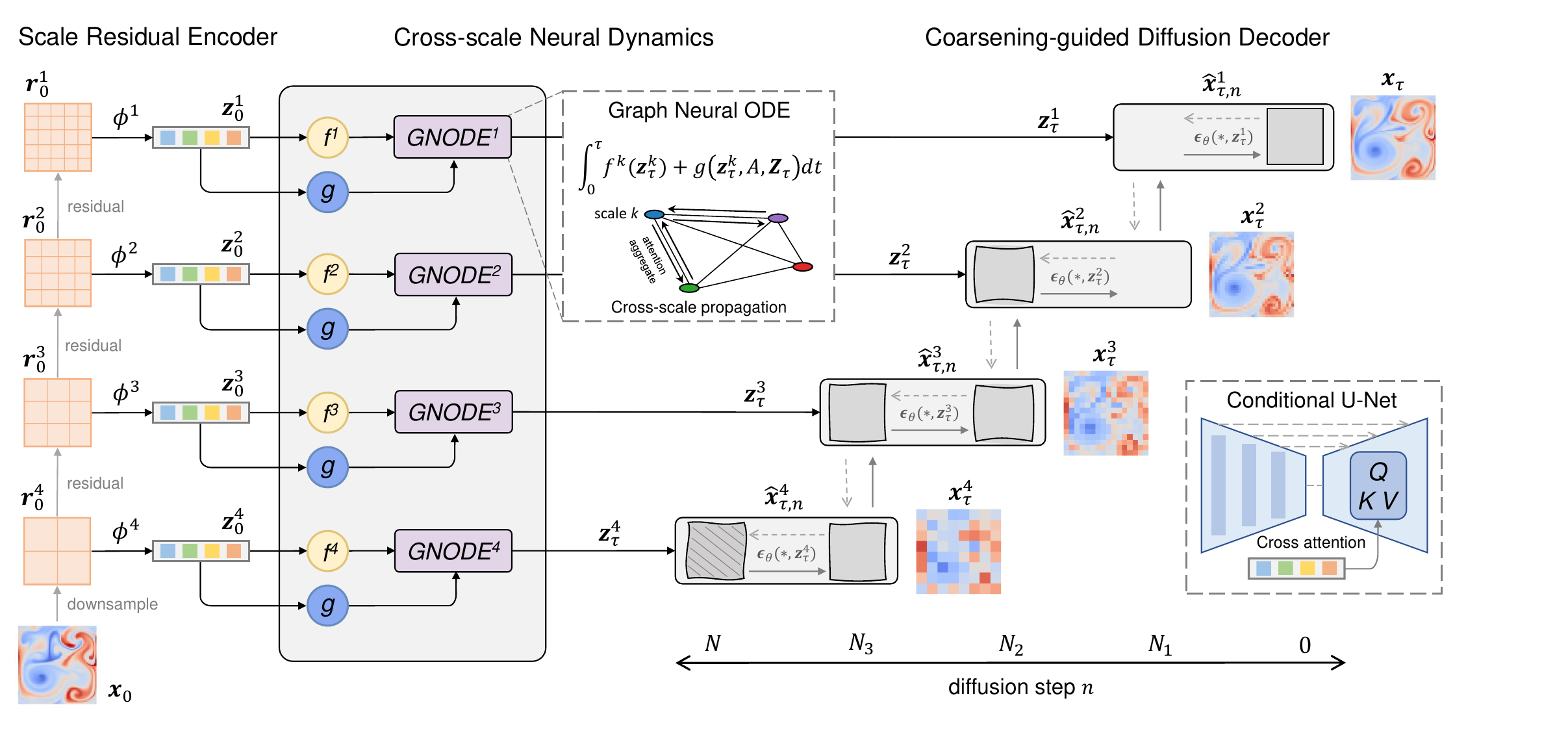}
    \caption{Overall framework of MDPNet.}
    \label{fig:framework}
\end{figure*}

In this section, we propose Multiscale Diffusion Prediction Network (MDPNet) to predict the dynamics of complex systems via encoding and predicting the latent dynamics of multiscale representations.  
According to the challenges mentioned above, we first propose a multiscale diffusion autoencoder, where a residual encoder and a diffusion decoder collaboratively capture the latent space of multiscale dynamics in complex systems. Then, we design a graph neural ordinary differential equation (GNODE) module to model scale-specific dynamics and cross-scale propagation for predicting latent dynamics.
The overall framework is illustrated in Figure~\ref{fig:framework}.

\subsection{Multiscale Diffusion Autoencoder}

To model the latent space of intrinsic dynamics, we first design a diffusion autoencoder that explicitly leverages the multiscale structure of complex systems. The diffusion decoder employs a diffusion model $\psi$ to fit the conditional distribution of complex spatiotemporal patterns $p(\bm{x}_\tau | \bm{z}_\tau)=\psi(\bm{z}_\tau)$, where the conditions $\bm{z}_\tau$ incorporate multiscale information provided by the encoder $\phi$. The encoder $\phi$ and decoder $\psi$ work collaboratively to represent spatiotemporal information across different scales, thereby uncovering the latent space where the intrinsic dynamics reside. In the following, we introduce the detailed design of the autoencoder.

\subsubsection{Multiscale Residual Encoder}

Here, we introduce a residual encoding method to decompose system state information across different scales. Given a total of $K$ scales, we decompose the observed state $\bm{x}_\tau$ into $\{\bm{x}^k_\tau\}^K_{k=1}$, where $k$ represents the scale level (with larger values corresponding to coarser granularity). Considering the directed influence from coarse to fine scales~\cite{lee2020coarse, vlachas2022multiscale}, we first extract the coarsest-scale feature as $\bm{x}^K_\tau=\mathcal{Q}(\bm{x}_\tau, K)$, where $\mathcal{Q}(*,k): \mathbb{R}^{H \times W} \rightarrow \mathbb{R}^{H \times W}$ denotes a downsampling operator by a factor of $k$ followed by interpolation back to the original resolution. Consequently, the residual $\bm{x}_\tau-\bm{x}^K_\tau$ captures the high-frequency information lost during coarse-scale encoding.

We then extract the $(K-1)$-scale features from the residual as $\bm{r}^{K-1}_\tau=\mathcal{Q}(\bm{x}_\tau-\bm{x}^K_\tau, K-1)$, and update the next-level residual as $\bm{x}_\tau-\bm{x}^K_\tau-\bm{r}^{K-1}_\tau$.
Let $\bm{r}^K_\tau=\bm{x}^K_\tau$, then the general formula for the residual at scale $k$ is given by 
\begin{equation}
    \bm{r}^k_\tau= \mathcal{Q}(\bm{x}_\tau-\sum^{K}_{i=k+1}\bm{r}^i_\tau, k),
\end{equation}
which naturally decomposes the information from coarse to fine scales.
The coarse-grained state at scale $k$ is obtained by accumulating the preceding residuals as 
\begin{equation}\label{equ:coarsen}
    \bm{x}^k_\tau=\sum^{K}_{i=k+1}\bm{r}^i_\tau.
\end{equation}
% While downsampling inevitably leads to information loss, our multiscale residual encoding ensures that the total loss of information decreases exponentially with scale number $k$.
Whereas traditional downsampling methods inevitably lead to information loss, our multiscale residual encoding ensures that information loss decreases exponentially with scale number $k$ and remains theoretically lossless when $k=1$ (i.e., $\bm{x}_\tau=\bm{x}^1_\tau$).
The similar idea has been validated in latest image representation studies~\cite{lee2022autoregressive, tian2024visual}.

Finally, we encode the features at each scale to obtain the multiscale representation of the system state, $\bm{z}^k_\tau=\phi^k(\bm{r}^k_\tau)$, where $\bm{z}^k_\tau$ encapsulates the latent dynamical information at scale $k$. Instead of independently training a separate encoder for each scale, we adopt a scale-aware encoder $\phi_\theta: \mathbb{R}^{C \times H \times W} \rightarrow \mathbb{R}^d$ with shared parameters to map the features at scale $k$ as $\bm{z}^k_\tau=\phi^k_\theta(\bm{r}^k_\tau)=\phi_\theta(\bm{r}^k_\tau, Embedding(k))$, where $Embedding(*)$ is a trainable scale embedding (inspired by positional encoding in Transformers~\cite{waswani2017attention}). This approach simultaneously preserves both feature details and scale-level information, resulting in multiscale $d$-dimensional latent vectors $\bm{Z}_\tau=\{\bm{z}^k_\tau\}^K_{k=1}$.

\subsubsection{Coarsening-guided Diffusion Decoder}\label{sec:decoder}

We integrate the coarse-to-fine multiscale reconstruction task with the diffusion process and propose a coarsening-guided diffusion decoder $\psi_\theta$. Building upon the reverse diffusion process introduced in Sec.~\ref{sec:pre_diff}, we sequentially allocate the total $N$ diffusion steps to each scale, yielding the schedule: $\{N_k\}^{K-1}_{k=1}$. The diffusion stage corresponding to the $k$-th scale spans from $N_{k}$ to $N_{k+1}$, during which the noise network estimate noise as $\bm{\epsilon}_\theta(\bm{x}_{\tau,n},n,\bm{z}_k)$, where $\bm{z}_k$ serves as conditional input. Next, we provide a detailed design of the forward and reverse processes of the coarsening-guided diffusion decoder. To avoid ambiguity, we use $\tau$ as the subscript for dynamical time steps and $n$ as the subscript for diffusion steps.

In the forward process, we propose a multi-stage noise scheduling strategy to integrate the coarse-graining process of complex systems, as described in Equation~\ref{equ:coarsen}, with the diffusion noise injection process.
Specifically, the noise scheduling at scale $k$ is defined as 
\begin{equation}
\alpha^k_n =
\begin{cases}
    1, & \text{if } n < N^k, \\
    \alpha_n, & \text{if } N^k \leq n < N^{k+1}.
\end{cases}
\end{equation}
where no noise is added before reaching the $k$-th stage. Instead, the process applies coarse-graining at scale $k$, 
\begin{equation}
    \bm{x}_{\tau,n}=\sqrt{\overline{\alpha}^k_n}\bm{x}^k_{\tau,0}+\sqrt{1-\overline{\alpha}^k_n}\bm{\epsilon}.
\end{equation}
This approach replaces the early stages ($n < N^k$) of the original diffusion noise injection process with the corresponding coarse-grained transformation, implicitly guiding the diffusion model to progressively perturb fine-grained distribution features in a coarse-grained manner during the early noise injection stages~\cite{fan2024mg}.

In the reverse process, we use the multiscale features $\bm{Z}_\tau$ as conditional information and sequentially guide the diffusion model's sampling in reverse order according to the schedule $\{N_k\}^{K-1}_{k=1}$. Specifically, during the diffusion stage of scale $k$, the noise network receives $\bm{z}^k_\tau$ as a conditional input to predict the noise as
\begin{equation}
     \bm{\epsilon}_n = \bm{\epsilon}_\theta(\bm{x}^k_{\tau,n},n,\bm{z}^k_\tau) = \bm{\epsilon}_\theta(\bm{x}^k_{\tau,n},n,\phi^k_\theta(\bm{r}^k_\tau)).
\end{equation}
As the reverse process progresses, the diffusion sample $\bm{\hat{x}}^k_{\tau,n}$ gradually aggregates information from coarse to fine scales, similar to the residual reconstruction process described in Equation~\ref{equ:coarsen}. However, instead of using the traditional decoder to reconstruct the residual $\bm{r}^k_\tau$ from $\bm{z}^k_\tau$, we use $\bm{z}^k_\tau$ as a condition term to guide the denoising direction of the diffusion model.
The former follows a conventional self-supervised reconstruction objective, requiring the latent vector $\bm{z}^k_\tau$ to preserve all features. In contrast, our approach aims to guide high-quality denoising, allowing the encoder, in collaboration with the diffusion decoder, to encode only multiscale conditional information. 
Within this paradigm, the encoder extracts multiscale structural information to guide the denoising process, while the diffusion model captures and refines the spatiotemporal patterns and textures of the data distribution~\cite{preechakul2022diffusion}.
Together, they form a novel autoencoder that shapes the low-dimensional latent space of intrinsic dynamics.

\subsection{Cross-scale Neural Dynamics}

We model multiscale dynamics $\bm{p}_\theta(\bm{z}^k_\tau | \bm{z}^k_0)$ in the low-dimensional latent space obtained in the previous step.
Considering cross-scale interactions, we extend the prediction of dynamics at each scale to a full-scale conditional probability $\bm{p}_\theta(\bm{z}^k_\tau | \bm{Z}_\tau)$. For a given scale number $K$, we construct a fully connected topology $A$ representing the interaction network among scales (i.e., nodes) and model the dynamics prediction at each scale as a co-evolution problem of node states on the graph.

We employ graph neural ordinary differential equations~\cite{zang2020neural} to model multiscale dynamics. The ODE function is defined as 
\begin{equation}
    \frac{d\bm{z}^k_\tau}{dt} = f^k(\bm{z}^k_\tau) + g(\bm{z}^k_\tau, A, \bm{Z}_\tau),
\end{equation}
where the self-dynamics $f^k$ captures scale-specific dynamics, and the interaction term $g$ models cross-scale information propagation.
We parameterize the self-dynamics as a scale-aware neural network $f^k(\bm{z}^k)=\xi_\theta(\bm{z}^k, Embedding(k))$, enabling parameter sharing to improve computational efficiency. For the interaction term, we employ a graph attention model~\cite{velivckovic2017graph} to automatically learn system-specific cross-scale information propagation~\cite{weinan2011principles}. Finally, the evolution trajectory of multiscale dynamics can be solved as an initial value problem 
\begin{equation}
    \bm{\hat{z}}^k_\tau = \bm{z}^k_0 + \int^\tau_0 f^k(\bm{z}^k) + g(\bm{z}^k_t, A, \bm{Z}_t) dt
\end{equation}
using any ODE solver. This allows us to predict the system state at arbitrary continuous time point $\tau$.

\subsection{Training}

The training process of the model consists of two stages: pretraining and end-to-end training. In the pretraining stage, the predictor is frozen, and only the autoencoder module is trained. During this stage, the encoded latent space undergoes rapid adjustment and gradually converges near an optimal point. The multiscale residual encoder and coarsening-guided diffusion decoder jointly update parameters to minimize the noise estimation error 
\begin{equation}
    L_{latent}=\mathbb{E}_{n,\bm{\epsilon}_n,\bm{x}_0}||\bm{\epsilon}_n-\bm{\epsilon}_\theta(\bm{x}_{\tau,n}, n, \bm{z}_\tau)||^2.
\end{equation}

In the end-to-end training stage, the predictor and autoencoder are trained jointly, and the diffusion decoder receives future predictions $\bm{\hat{z}}^k_\tau$ from the predictor as conditional inputs. The prediction loss is computed as the mean squared error of future latent vectors, given by $L_{pred}=\mathbb{E}_{k,\bm{x}_\tau}||\bm{z}^k_\tau -  \bm{\hat{z}}^k_\tau||^2$, guiding the training of the entire model with the denoising loss.

\section{Experiments}

In this section, we present the extensive evaluation results of MDPNet. We analyze the prediction performance of all models on four representative complex systems, followed by additional experiments to analyze MDPNet’s hyperparameter sensitivity, robustness, interpretability, generalization, ablation study, and computational cost.

\subsection{Datasets}
We consider the following four classical partial differential equation systems with complex spatiotemporal patterns:
\begin{itemize}
    \item \textbf{Lambda–Omega equation} (LO)~\cite{champion2019data} describes a classic reaction-diffusion system with two interacting components, capturing complex spatiotemporal dynamics and pattern formation.
    \item \textbf{Brusselator equation} (Bruss)~\cite{prigogine1967symmetry}'s long-term dynamics converge to a limit cycle, indicating that after the initial transient phase, the system’s trajectory will approach a specific periodic orbit.
    \item \textbf{Gray-Scott euqation} (GS)~\cite{rao2023encoding} models the self-organizing process of chemical substances diffusing and reacting in space, exhibiting a rich variety of dynamic behaviors and distinct pattern formations.
    \item \textbf{Incompressible Navier-Stokes equation} (NS)~\cite{takamoto2022pdebench} is the incompressible version of the fluid dynamics equations, describing subsonic flows and wave propagation in systems ranging from hydrodynamics to weather prediction.
\end{itemize}
We simulate trajectories from 100 different initial conditions as the training set, with 50 for testing (except for the NS system). All system trajectory lengths are unified to 100 steps (50 steps for the NS system). Details on data generation and preprocessing can be found in Appendix~\ref{app:exper_setup}.

\subsection{Baselines}
For all the datasets, we compare with the following representative methods.
\begin{itemize}
    \item \textbf{FNO} \cite{lifourier} leverages Fourier transforms to learn and approximate solution operators for partial differential equations directly in the frequency domain.
    \item \textbf{ConvLSTM} \cite{shi2015convolutional} incorporates convolutional operations into LSTM’s input-to-state and state-to-state transitions.
    \item \textbf{Neural ODE} \cite{chen2018neural} parameterizes continuous-time hidden dynamics with a neural network and solves them using an ODE solver.
    \item \textbf{DeepONet} \cite{lu2021learning} learns nonlinear operators using a dual-network architecture, where a branch network encodes input functions and a trunk network encodes locations.
    \item \textbf{UNet} \cite{ronneberger2015u}: employs an encoder-decoder architecture with skip connections to capture both global context and fine-grained details.
    % \item \textbf{DYffusion} \cite{ruhling2024dyffusion}: reimagines diffusion process by coupling temporal interpolation with forecasting, imposing a dynamics-informed inductive bias for spatiotemporal prediction.
    \item \textbf{FNO-coarsen} \cite{lifourier} extends FNO by conducting predictions in a downsampled representation of the spatial resolution.
    \item \textbf{AE-LSTM} \cite{vlachas2022multiscale} utilizes an autoencoder to project data into a latent space, where an LSTM models temporal dynamics for prediction.
    \item \textbf{Latent ODE} \cite{chen2018neural} encodes data into a latent space and models its continuous-time dynamics using a neural ODE solver.
    \item \textbf{L-DeepONet} \cite{kontolati2024learning} reformulates DeepONet in a latent space, using an autoencoder to map high-dimensional data to a lower-dimensional representation.
    \item \textbf{G-LED} \cite{gao2024generative} evolves latent dynamics with autoregressive attention and reconstructs high-dimensional states using Bayesian diffusion model.
\end{itemize}
We categorize these baselines into two types based on the state space where the dynamics prediction is executed: observed-space and latent-space, as shown in Table~\ref{tab:main_result}. In latent-space methods, predictions occur in the latent space after encoding or downsampling.

\subsection{Setup}

For all models, we split the dataset into an 8:2 training-to-validation ratio. During training, the loss is calculated based on the prediction results at 5-step intervals, while during testing, we autoregressively predict the entire trajectory starting from the initial conditions. 
We evaluate the predicted trajectory against the true trajectory using normalized mean squared error (NMSE)~\cite{ohana2024well} and structural similarity index (SSIM)~\cite{hore2010image} to assess the error at each grid point and the global structural similarity. NMSE is sensitive to numerical anomalies, while SSIM captures the structural and textural patterns of spatiotemporal dynamics, making the two metrics complementary and effective for evaluating prediction quality (details in the Appendix~\ref{app:exper_setup}).
MDPNet and G-LED use 1,000 diffusion steps, with the scale schedule $\{N_k\}^{K-1}_{k=1}$ for MDPNet being uniformly distributed. Unless otherw

In contrast, MDPNet, leveraging its novel multiscale diffusion autoencoder, effectively captures a meaningful latent space across all systems. We further analyze the advantages of the multiscale diffusion autoencoder in Sec.~\ref{sec:autoencoder} and the cross-scale neural dynamics module in Sec.~\ref{sec:ablation}.

\begin{table*}[!ht]
\renewcommand{\arraystretch}{1.1}
\caption{Average performance of the trajectories predicted from different initial conditions with standard deviation from 10 runs. The best results are highlighted in bold, and the suboptimal results are emphasized with an underline.}
\centering
\resizebox{\textwidth}{!}{%
\begin{tabular}{ll|c|c|c|c|c|c|c|c}
\hline \hline
 & \multirow{2}{*}{Methods} & \multicolumn{2}{|c}{Lambda-Omega} & \multicolumn{2}{|c}{Brusselator} & \multicolumn{2}{|c}{Gray-Scott} & \multicolumn{2}{|c}{Navier-Stokes} \\
 & & \multicolumn{1}{|c}{NMSE  $\times 10^{-2} \downarrow$} & \multicolumn{1}{c}{SSIM  $\times 10^{-1} \uparrow$} & \multicolumn{1}{|c}{NMSE $\times 10^{-2} \downarrow$} & \multicolumn{1}{c}{SSIM $\times 10^{-1} \uparrow$} & \multicolumn{1}{|c}{NMSE $\times 10^{-2} \downarrow$} & \multicolumn{1}{c}{SSIM $\times 10^{-1} \uparrow$} & \multicolumn{1}{|c}{NMSE $\times 10^{-2} \downarrow$} & \multicolumn{1}{c}{SSIM $\times 10^{-1} \uparrow$} \\

\midrule
\multirow{5}{*}{\rotatebox{90}{Observed-space}}
 & ConvLSTM & $ 7.074 \pm 1.580 $ & $8.376\pm 0.134$ & $\underline{0.886\pm 0.070}$ & $\underline{9.287\pm 0.088}$ & $3.282\pm0.171$ & $7.944\pm0.342$ & $\underline{1.996\pm0.003}$ & $\underline{7.580\pm0.005}$ \\
 & Neural ODE & $18.919\pm 1.259$ & $4.253\pm 0.050$ & $10.513 \pm 1.064$ & $4.675\pm 1.051$ & $11.836\pm 1.139$ & $4.181\pm 1.146$ & $2.575\pm 0.001$ & $6.452\pm0.003$ \\
 & DeepONet & $19.775\pm11.202$ & $4.862\pm1.276$ & $13.189\pm 3.671$ & $6.527\pm0.339$ & $13.590\pm4.715$ & $2.684\pm1.331$ & $15.776\pm5.063$ & $3.102\pm0.849$ \\
 & FNO & $3.768\pm0.287$ & $8.704\pm0.200$ & $5.349\pm 0.428$ & $7.453\pm 0.751$ & $5.916\pm 0.426$ & $6.567\pm 0.700$ & $2.416\pm 0.010$ & $7.137\pm 0.135$ \\
 & UNet & $16.290\pm5.450$ & $5.773\pm0.807$ & $16.805\pm0.396$ & $6.034\pm0.089$ & $4.380\pm0.027$ & $6.173\pm0.022$ & $2.291\pm0.273$ & $7.381\pm0.118$ \\
 % & DYffusion & $0.xxx\pm0.xxx$ & $0.xxx\pm0.xxx$ & $0.xxx\pm0.xxx$ & $0.xxx\pm0.xxx$ & $0.xxx\pm0.xxx$ & $0.xxx\pm0.xxx$ & $0.xxx\pm0.xxx$ & $0.xxx\pm0.xxx$ \\
 
\midrule
\multirow{6}{*}{\rotatebox{90}{Latent-space}} 
 & AE-LSTM & $6.214\pm 0.706$ & $8.043\pm 0.195$ & $7.496\pm 0.634$ & $7.782\pm 0.286$ & $6.240\pm 0.518$ & $4.609\pm 0.729$ & $2.038\pm 0.003$ & $ 7.369\pm 0.006$ \\
 & Latent ODE & $34.539\pm 0.752$ & $3.777\pm 0.062$ & $10.249\pm 0.579$ & $7.011\pm 0.216$ & $15.009\pm 0.852$ & $2.049\pm 0.171$ & $2.259\pm 0.003$ & $ 7.378\pm 0.006$ \\
 & L-DeepONet & $9.299\pm 6.999$ & $6.869\pm 1.021$ & $6.087\pm 0.250$ & $7.188\pm 0.294$ & $\underline{3.117\pm 0.249}$ & $\underline{8.016\pm 0.345}$ & $2.190\pm 0.004$ & $ 7.397\pm 0.007$ \\
 &  FNO-coarse & $6.891\pm 0.963$ & $7.618\pm 0.270$ & $9.589\pm 1.438$ & $7.212\pm 0.784$ & $4.378\pm 0.195$ & $6.852\pm 0.526$ & $2.410\pm 0.009$ & $ 7.074\pm 0.129$ \\
 & G-LED & $\underline{1.307\pm0.503}$ & $\underline{9.140\pm0.203}$ & $10.534\pm0.035$ & $7.987\pm0.049$ & $10.117\pm0.069$ & $4.190\pm0.058$ & $2.689\pm0.285$ & $7.071\pm0.106$ \\
 & Ours & $\mathbf{1.145\pm0.513}$ & $\mathbf{9.276\pm0.172}$ & $\mathbf{0.044\pm0.029}$ & $\mathbf{9.900 \pm0.020}$ & $\mathbf{1.144\pm0.071}$ & $\mathbf{8.187\pm0.128}$ & $\mathbf{1.154\pm0.476}$ & $\mathbf{8.371\pm0.382}$ \\
 & PROMOTION & 12.40\% & 1.49\% & 95.03\% & 6.60\% & 63.30\% & 2.13\% & 42.18\% & 10.44\% \\

 \hline \hline
\end{tabular}%
}
\label{tab:main_result}
\end{table*}

\subsection{Sensitivity Analysis}

In this section, we analyze the impact of two important hyperparameters, namely the latent dimension and the scale number, on the prediction performance of MDPNet.

\subsubsection{Latent Dimension}

The $latent \ dimension \ d$ is the encoding dimension for each scale, which affects the information capacity of the latent vector. We compare the performance of two other latent space-based prediction models, AE-LSTM and G-LED, on the Bruss and NS systems, using the same encoding dimension to compare the performance of different algorithms. Considering that MDPNet encodes information across multiple scales, we set the encoding dimension for the comparison models to be $K$ (i.e., $scale \ number$) times the latent dimension for fairness.

\begin{figure}[!ht]
\centering
\subfigure[Bruss]{
\includegraphics[width=0.485\linewidth]{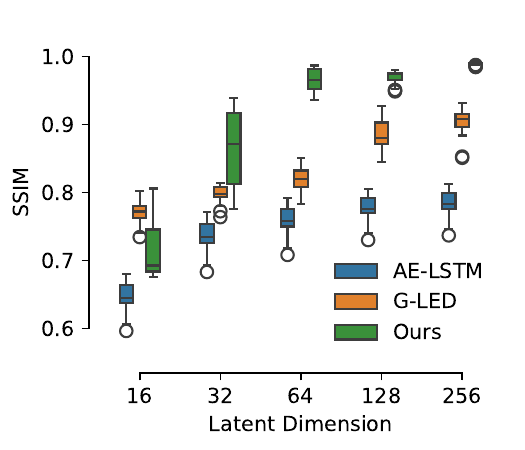}}
\subfigure[NS]{
\includegraphics[width=0.485\linewidth]{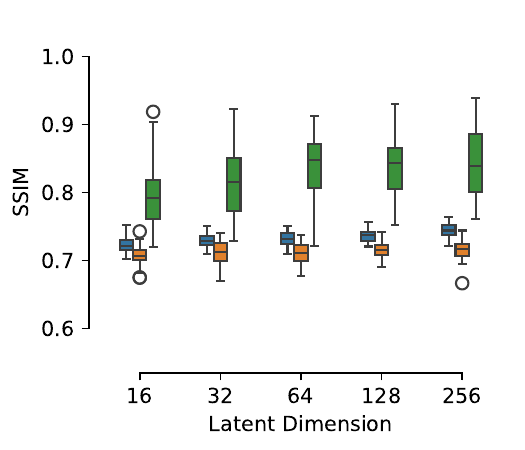}}
\caption{SSIM as a function of latent dimension for Bruss and NS systems.}
\label{fig:dimension}
\end{figure}

Figure~\ref{fig:dimension} shows that the performance of our MDPNet converges at 64 dimensions, with an upper bound on accuracy significantly higher than that of baselines. Although all are latent space prediction algorithms, MDPNet is more accurate than the single-scale baselines by modeling the characteristics and interactions of dynamics across multiple scales. 
On the Bruss system, when the latent dimension is below 64, it becomes the performance bottleneck for all algorithms. MDPNet achieves near-optimal performance at 64 dimensions, while G-LED requires 128 dimensions to converge (and its accuracy is lower than that of MDPNet). 
On one hand, as described in Sec.~\ref{sec:decoder}, MDPNet optimizes the latent vector $\bm{z}^k_\tau$ for guiding diffusion denoising rather than lossless reconstruction. Consequently, it imposes lower requirements on the encoding dimension (i.e., capacity) compared to traditional autoencoders, allowing MDPNet to outperform them in most cases with the same encoding dimension.
On the other hand, compared to G-LED, MDPNet’s multiscale residual encoder mitigates information loss during downsampling and decouples multiscale dynamics to enhance prediction accuracy. As a result, it achieves a significantly higher performance upper bound.

\subsubsection{Scale Number}

The $scale \ number \ K$ is the number of scales encoded, which controls the granularity with which MDPNet perceives the system state. Considering that the number of diffuse steps allocated to different scales may affect the prediction results, we fix the allocation of 200 diffuse steps for each scale. We test 1 to 5 scales, with a total diffuse step count ranging from 200 to 1,000. Additionally, we conduct a control experiment with a single scale but varying the number of diffuse steps from 200 to 1,000 to eliminate the impact of total diffuse steps.

Table~\ref{tab:scale_num} shows the experimental results on the NS system. Increasing the diffuse steps and $scale \ number$ both improve predictions in the early stages. However, for the same diffuse steps, the single-scale MDPNet consistently performs worse than the multi-scale version. This indicates that a denoising process guided by a coarse-to-fine approach can effectively improve the reconstruction quality of diffusion. Furthermore, when the $scale \ number$ exceeds 3, the prediction performance no longer improves and slightly declines. This is because, once the observational granularity exceeds a certain threshold, residual information at coarser scales provides limited additional information gain and may increase the risk of overfitting (details in the Appendix Sec.~\ref{Sec:apx_scale_residual}). Therefore, it is generally recommended to set the $scale \ number$ to 3 and fine-tune it based on the specific characteristics of the system.

\begin{table}[!ht]
\renewcommand{\arraystretch}{1.1}
\caption{Prediction performance as functions of diffuse steps and scale numbers for NS system.}
\centering
\resizebox{\linewidth}{!}{%
\begin{tabular}{p{0.8cm}|c|c|p{0.7cm}|c|c}
\hline
 Diffuse Steps &  NMSE  $\times 10^{-2} \downarrow$ & SSIM  $\times 10^{-1} \uparrow$ & Scale Num & NMSE $\times 10^{-2} \downarrow$ & SSIM $\times 10^{-1} \uparrow$ \\
 
 \midrule
$200$ & $4.236\pm1.415$ & $6.037\pm0.897$ & 1 & $4.236\pm1.415$ & $6.037\pm0.897$ \\
$400$ & $2.071\pm0.667$ & $7.713\pm0.354$ & 2 & $1.910\pm0.489$ & $7.754\pm0.321$ \\
$600$ & $2.067\pm0.989$ & $7.865\pm0.566$ & 3 & $\mathbf{1.188\pm0.498}$ & $\mathbf{8.313\pm0.396}$ \\
$800$ & $1.673\pm0.691$ & $8.033\pm0.477$ & 4 & $1.269\pm0.611$ & $8.212\pm0.446$ \\
$1000$ & $\mathbf{1.641\pm0.638}$ & $\mathbf{8.081\pm0.408}$ & 5 & $1.256\pm0.640$ & $8.295\pm0.373$ \\

 \hline
\end{tabular}%
}
\label{tab:scale_num}
\end{table}

\subsection{Robustness Analysis}

\begin{figure}[!ht]
\centering
\subfigure[Noise]{
\includegraphics[width=0.485\linewidth]{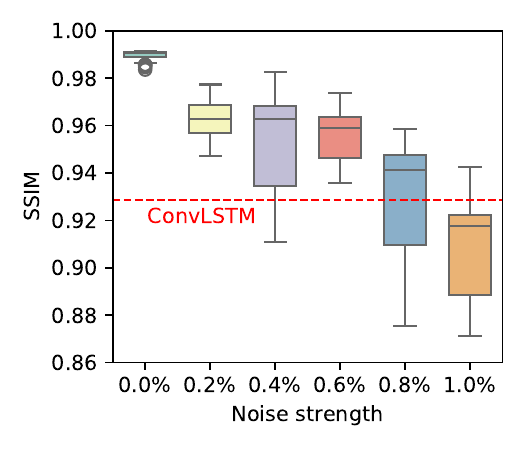}}
\subfigure[Data size]{
\includegraphics[width=0.485\linewidth]{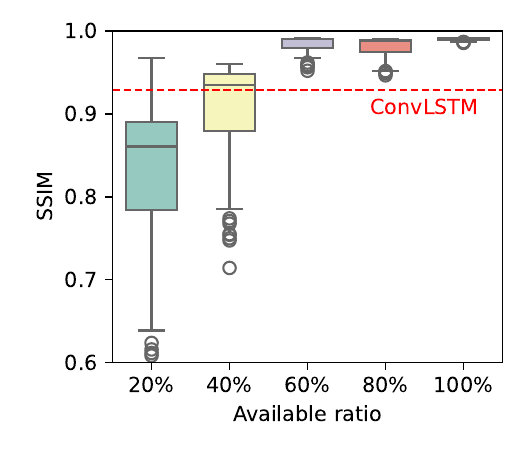}}
\caption{SSIM distribution as a function of (a) noise strength and (b) available training ratio for Bruss system.}
\label{fig:robustness}
\end{figure}

We use the Bruss system as an example to evaluate MDPNet's robustness under noisy and data-scarce conditions. During training, we introduce Gaussian noise of varying relative strengths to observe its impact on MDPNet. The results show that MDPNet is robust to data noise (Figure~\ref{fig:robustness}a) and outperforms most baseline algorithms trained on noise-free data, even in the presence of noise. Additionally, we reduce the number of trajectories in the training set to examine MDPNet's performance fluctuations when available data is limited, as shown in Figure~\ref{fig:robustness}b. Even with only 60\% of the training data, MDPNet outperforms the optimal baseline that uses the full dataset. This result highlights MDPNet's ability to effectively incorporate multi-scale information enhances its efficiency in utilizing available data, making it robust even when data is limited.

\subsection{Interpretability Analysis}\label{sec:autoencoder}

\begin{figure}[!ht]
\centering
\subfigure[Scale 1]{
\includegraphics[width=0.31\linewidth]{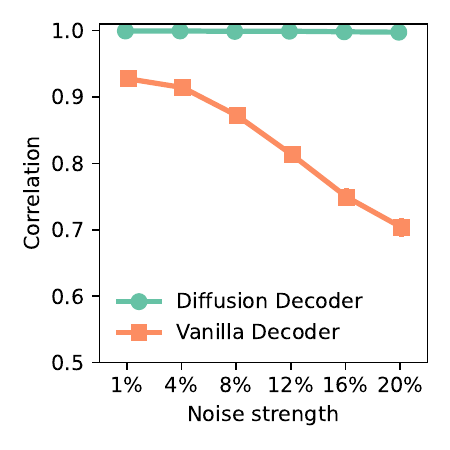}}
\subfigure[Scale 2]{
\includegraphics[width=0.31\linewidth]{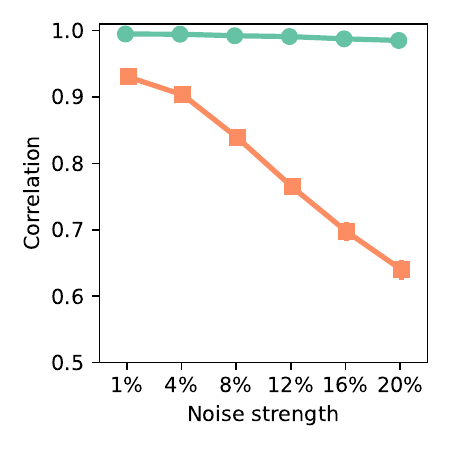}}
\subfigure[Scale 3]{
\includegraphics[width=0.31\linewidth]{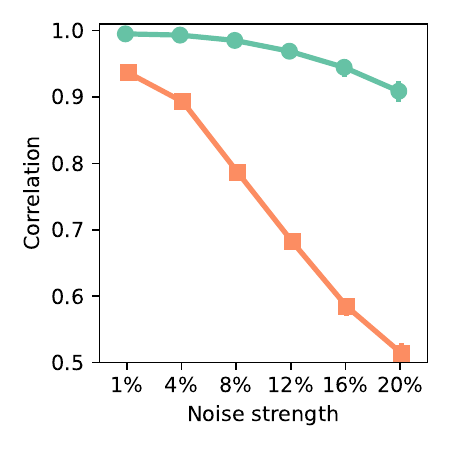}}
\caption{Pearson correlation coefficient of different decoders as a function of noise intensity at different scales.}
\label{fig:interpretability}
\end{figure}

Using a three-scale MDPNet on the GS system as an example, we explore the resilience of the diffusion decoder to information at different scales, thereby revealing the underlying mechanisms of MDPNet. Specifically, we inject Gaussian noise with varying relative strengths into the latent vector $z^k_\tau$ at each scale $k$ to simulate different levels of degradation in the corresponding latent vector and use the decoder to reconstruct the observed state $\bm{x}_\tau$. To control variables, we perturb only a single scale at a time. 
Then, we take this a step further by re-encoding $\bm{x}_\tau$ using the encoder to obtain $\bm{\overline{z}}^k_\tau$ By comparing the correlation between $\bm{z}^k_\tau$ and $\bm{\overline{z}}^k_\tau$, we quantitatively assess whether the encoder and decoder can collaboratively tolerate errors and mitigate the accumulation of long-term prediction errors in the latent space. 
As shown in Figure~\ref{fig:interpretability}, compared to a vanilla decoder, our coarsening-guided diffusion decoder exhibits strong resilience to perturbations at every scale. 
We guide the diffusion decoder to reconstruct the complex patterns of spatiotemporal distributions, rather than naively reconstructing residuals at different scales and summing them according to Equation~\ref{equ:coarsen}, which leads to accumulated errors.
This explains MDPNet’s consistently superior long-term prediction performance observed in Table~\ref{tab:main_result}.

\subsection{Generalization Analysis}

To assess the generalization of MDPNet to generate spatiotemporal patterns under unseen parameter control, we choose cylinder flow dynamics as the experimental subject. Cylinder flow describes the process where a fluid forms a Kármán vortex street after passing a cylindrical inlet \cite{vlachas2022multiscale}, whose velocity field is governed by the Navier-Stokes equations. The Reynolds number of the system affects the size and frequency of the vortices, with moderate Reynolds numbers inducing periodic arrangement of vortex streets in the flow. We uniformly collect 50 points of Reynolds numbers in the range from 100 to 500, and simulate these 50 evolutionary trajectories as the training set using the lattice Boltzmann method (details in the Appendix~\ref{app:exper_setup}). Subsequently, we divide the test set into two groups: in-distribution and out-of-distribution. In-distribution test trajectories correspond to 10 uniformly sampled Reynolds numbers in the range from 100 to 500, while out-of-distribution test trajectories correspond to the range from 500 to 1,000.

\begin{figure}[!ht]
\centering
\subfigure[In-distribution]{
\includegraphics[width=0.485\linewidth]{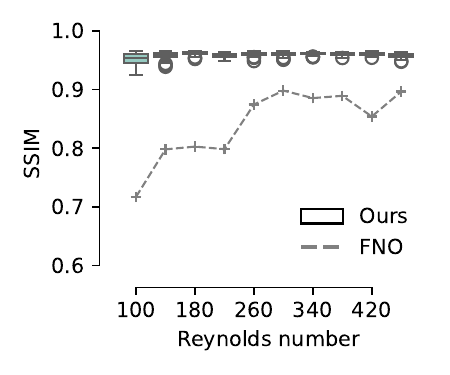}}
\subfigure[Out-of-distribution]{
\includegraphics[width=0.485\linewidth]{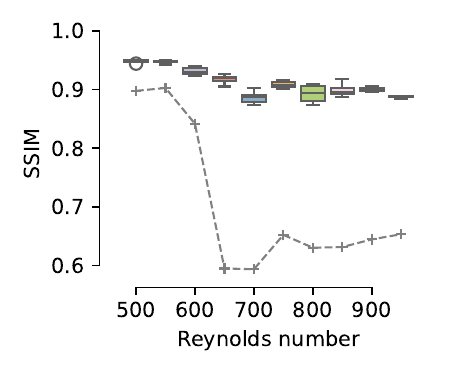}}
\caption{SSIM as a function of Reynolds number for cylinder flow system.}
\label{fig:generalization}
\end{figure}

\begin{figure}[!ht]
\centering
\subfigure[In-distribution]{
\includegraphics[width=0.485\linewidth]{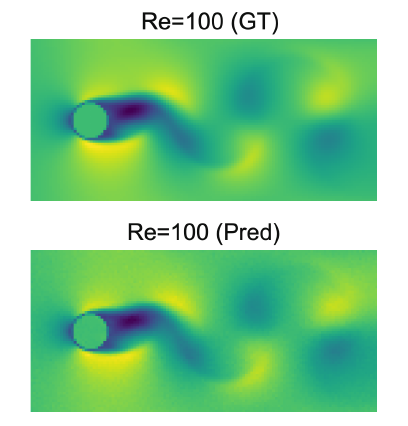}}
\subfigure[Out-of-distribution]{
\includegraphics[width=0.485\linewidth]{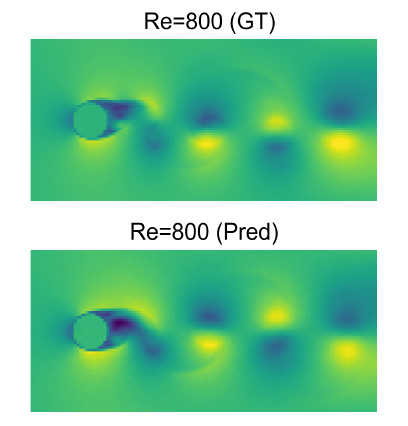}}
\caption{Snapshots of different Reynolds numbers for cylinder flow system.}
\label{fig:generalization_snapshot}
\end{figure}

The predictive performance of MDPNet consistently outperforms the classic baseline (Figure~\ref{fig:generalization}). As the Reynolds number exceeds the range seen during training, there is a slight decline in the prediction accuracy of MDPNet, whereas the baseline algorithm shows a significant deterioration. We report prediction snapshots at both high and low Reynolds numbers in Figure~\ref{fig:generalization_snapshot}. In in-distribution scenarios (even when specific Reynolds values were not seen during training), MDPNet is able to accurately predict the turbulent patterns and waveforms on the exterior of the cylinder. Even as the Reynolds number moves outside the training distribution, MDPNet still accurately predicts the number, shape, and relative position of the vortices. These results validate the generalization capability of MDPNet.

\subsection{Ablation Study}\label{sec:ablation}

Here, we conduct an ablation analysis of the multiscale neural dynamics module. We evaluate two ablated versions: (i) disabling the interaction term in the graph neural ODE, allowing each scale to make independent predictions (reducing to a standard Neural ODE); (ii) replacing the predictor with a vanilla alternative (e.g., LSTM) to assess the fault tolerance of our multiscale diffusion autoencoder. The results on LO and Bruss systems are presented in Table~\ref{tab:ablation}.
Disabling the cross-scale interaction term leads to a significant decline in MDPNet's prediction performance, validating the importance of decoupling and modeling the co-evolution of complex systems across scales. Moreover, even when replacing the predictor with a vanilla alternative, MDPNet still outperforms more than half of the baselines. This demonstrates that the structured encoding of multiscale information effectively preserves essential dynamical features, even with a simplified predictor.

\begin{table}[!ht]
\renewcommand{\arraystretch}{1.1}
\caption{Ablation study on LO and Bruss systems.}
\centering
\resizebox{\linewidth}{!}{%
\begin{tabular}{l|c|c|c|c}
\hline
 \multirow{2}{*}{Ablated versions} & \multicolumn{2}{|c}{Lambda-Omega} & \multicolumn{2}{|c}{Brusselator} \\
 & \multicolumn{1}{|c}{NMSE  $\times 10^{-2} \downarrow$} & \multicolumn{1}{c}{SSIM  $\times 10^{-1} \uparrow$} & \multicolumn{1}{|c}{NMSE $\times 10^{-2} \downarrow$} & \multicolumn{1}{c}{SSIM $\times 10^{-1} \uparrow$} \\
 \hline
 Ours \textsubscript{Neural ODE} & $4.851\pm0.545$ & $8.387\pm0.354$ & $7.403\pm0.542$ & $7.458\pm0.207$ \\
 Ours \textsubscript{LSTM} & $5.281\pm0.483$ & $8.029\pm0.627$ & $10.339\pm1.351$ & $7.267\pm0.462$ \\
 Ours & $\mathbf{1.145\pm0.513}$ & $\mathbf{9.276\pm0.172}$ & $\mathbf{0.044\pm0.029}$ & $\mathbf{9.900 \pm0.020}$ \\
 \hline
\end{tabular}%
}
\label{tab:ablation}
\end{table}

\subsection{Computational Cost}

In traditional high-fidelity numerical simulations, the entire process maintains full spatial resolution. Our method reduces the spatial dimension from $C \times H \times W$ to $K \times d$ dimensions. Taking the simulation of cylinder flow as an example, MDPNet reduces the original spatial dimensions from $2 \times 128 \times 64$ to $3 \times 128$, achieving over a 40-fold reduction.
Although the denoising process operates at the original spatial resolution, the number of denoising steps remains a fixed constant and does not increase with the prediction horizon. Consequently, the additional computational cost of MDPNet primarily comes from the prediction overhead in the low-dimensional latent space.

\begin{figure}[!ht]
    \centering
    \includegraphics[width=1\linewidth]{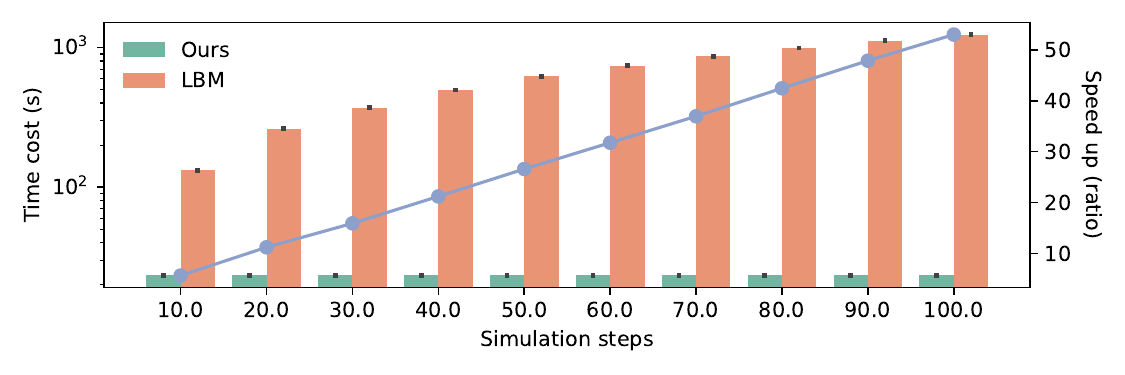}
    \caption{Time cost as a function of simulation steps for cylinder flow.}
    \label{fig:cy_time}
\end{figure}

We evaluate the computational cost of MDPNet by simulating the evolution $\bm{p}(\bm{x}_\tau | \bm{x}_0)$ from the initial state to time $\tau$ in the cylinder flow system, comparing it with lattice Boltzmann method (LBM)~\cite{vlachas2022multiscale}. As shown in Figure~\ref{fig:cy_time}, MDPNet with low-dimensional latent prediction exhibits a much slower increase in time cost with $\tau$ compared to traditional numerical methods. 
% Moreover, even in the most extreme case where diffusion reconstruction is required at every time step, the linear growth trend of MDPNet’s computational cost with $\tau$ is significantly slower than that of conventional numerical simulations (Appendix~\ref{app:exper_add}).

\section{Related Work}

\subsection{Discover Latent Space of Complex System}

A core challenge in complex system modeling is discovering the latent space where the intrinsic dynamics reside.
To accelerate the numerical solution of dynamical systems in rule-based domains, data-driven models~\cite{bar2019learning, lee2020coarse, vlachas2022multiscale, gao2024generative, wang2024p} have been proposed to identify coarse-scale PDEs that describe the evolution of macroscopic systems.
For unstructured data, renormalization group methods~\cite{garcia2018multiscale} aggregate node states to obtain a coarse-grained collective dynamics space.
Markov state encoding methods (e.g., VAMPnets~\cite{mardt2018vampnets}, T-IB~\cite{federici2023latent}, and NeuralMJP~\cite{seifner2023neural}) are designed to uncover the main metastable states of molecular kinetics.
Furthermore, physics-informed autoencoders have been developed to discover latent spaces governed by physical rules, such as linear operator~\cite{lusch2018deep}, nonlinear operator~\cite{kontolati2024learning} and the manifold of delayed embeddings~\cite{wu2024predicting}.
Compared to these methods, our aim is to discover a latent space that integrates the multiscale structure of complex systems, enabling more accurate predictions of system dynamics.

\subsection{Multiscale Modeling of Dynamics Prediction}

Complex systems consist of numerous interacting components, such as molecular particles in a chemical reaction or neurons in the brain~\cite{wang2024multi}. The nonlinear interactions and feedback mechanisms at the microscopic level give rise to emergent ordered structures at the macroscopic scale, motivating the exploration of self-organizing dynamical mechanisms~\cite{haken2006information, haken2012advanced} and cross-scale co-evolution from a multiscale perspective~\cite{peng2021multiscale, cash2006scale}.
\citet{bhatia2021machine} performs adaptive multiscale simulations of the interaction between RAS proteins and the plasma membrane. The simulation employs dynamic density functional theory at the macroscopic level and molecular dynamics simulations at the microscopic level.
\citet{vlachas2022multiscale} uses an autoencoder to extract the macroscopic state of a high-dimensional PDE system and alternately predicts dynamics at both the macroscopic and microscopic scales.
\citet{li2023learning} differentiates the fast and slow components in the overall dynamics based on changes in intrinsic dimensions, and uses Koopman operators and autoregressive models to predict these components separately.
These methods independently predict dynamics at individual scales at each time step, neglecting the propagation of information across scales.
\citet{wang2020deep} generates coarse-grained fluid dynamics data using a reduced-order model and then employs a neural network to model the correlation between the coarse data and the observed fine-scale data.
MultiScaleGNN~\cite{lino2021simulating} utilizes a U-Net-based graph neural network to implicitly capture multiscale information and propagates messages through edge connections.
In addition, numerous studies~\cite{gupta2022non, xiao2022coupled, gupta2021multiwavelet} leverage wavelet transforms for the analysis of multi-scale features.
Compared to these works, our model explicitly decouples multiscale representations while preserving cross-scale interactions.

\subsection{Diffusion Models for Dynamics Prediction}

The tremendous success of diffusion models in video generation~\cite{tumanyan2023plug, jin2024pyramidal} and time series modeling~\cite{fan2024mg, shen2024multi} has inspired a series of works in complex system dynamics prediction~\cite{li2025predicting2}.
\citet{shu2023physics} and \citet{li2024learning} have leveraged large-scale pre-trained diffusion models to reconstruct high-fidelity data from conventional low-fidelity samples or sparse measurement data. Known partial differential equations provide physical conditioning information for the denoising process, enhancing accuracy.
Building upon state-of-the-art diffusion models, \citet{li2024synthetic} propose a machine learning approach to generate single-particle trajectory data in high Reynolds number three-dimensional turbulence.
Similarly, \citet{lienen2023zero} treats turbulence simulation as a generative task, using a diffusion model to capture the distribution of turbulence induced by unseen objects and generate high-quality samples for downstream applications.
A recent study~\cite{bastek2024physics} introduces a first-principles-based loss term as physical knowledge to enhance the training process of the diffusion model, thus generating data samples that satisfy physical constraints.
\citet{ruhling2024dyffusion} aligns the temporal axis of spatiotemporal dynamics with diffusion’s process, replacing the noise injection with temporal interpolation and denoising with prediction, thereby embedding dynamical information into the diffusion process.
G-LED~\cite{gao2024generative} simply subsamples the system states as prediction targets, and uses predicted future frames as conditions for the diffusion model to reconstruct the high-fidelity original states.
In contrast to these works, we combine a scale residual encoder with diffusion to collaboratively discover the latent space of complex systems. This multiscale information gradually guides the denoising process of diffusion, thereby improving reconstruction quality.

\section{Conclusions}
In this paper, we design a Multiscale Diffusion Prediction Network (MDPNet) that leverages the inherent multiscale structure of complex systems to discover the latent space of intrinsic dynamics. 
By encoding multiscale conditions to guide the diffusion model in capturing spatiotemporal distributions, we extract latent vectors at different scales. 
We then employ an attention-based graph neural ordinary differential equation to model cross-scale interactions, enabling accurate predictions. 
Extensive experiments demonstrate that our model outperforms baselines in terms of accuracy, robustness, and generalization. 
Additionally, we analyze the effective gains from modeling cross-scale interactions on prediction performance.

In terms of limitations, this paper focuses on grid-based data. A direction for future work is to extend MDPNet's multiscale concept to spatiotemporal dynamics prediction on graph-structured data.

\begin{acks}
    This work is supported in part by the National Key Research and Development Program of China under 2024YFC3307603, and the National Natural Science Foundation of China under 62171260. 
\end{acks}

\bibliographystyle{ACM-Reference-Format}
\balance
\bibliography{reference}


\begin{thebibliography}{79}

%%% the Web address.

\ifx \showCODEN    \undefined \def \showCODEN     #1{\unskip}     \fi
\ifx \showISBNx    \undefined \def \showISBNx     #1{\unskip}     \fi
\ifx \showISBNxiii \undefined \def \showISBNxiii  #1{\unskip}     \fi
\ifx \showISSN     \undefined \def \showISSN      #1{\unskip}     \fi
\ifx \showLCCN     \undefined \def \showLCCN      #1{\unskip}     \fi
\ifx \shownote     \undefined \def \shownote      #1{#1}          \fi
\ifx \showarticletitle \undefined \def \showarticletitle #1{#1}   \fi
\ifx \showURL      \undefined \def \showURL       {\relax}        \fi
% invisible to TeX
\providecommand\bibfield[2]{#2}
\providecommand\bibinfo[2]{#2}
\providecommand\natexlab[1]{#1}
\providecommand\showeprint[2][]{arXiv:#2}

\bibitem[Bar-Sinai et~al\mbox{.}(2019)]%
        {bar2019learning}
\bibfield{author}{\bibinfo{person}{Yohai Bar-Sinai}, \bibinfo{person}{Stephan Hoyer}, \bibinfo{person}{Jason Hickey}, {and} \bibinfo{person}{Michael~P Brenner}.} \bibinfo{year}{2019}\natexlab{}.
\newblock \showarticletitle{Learning data-driven discretizations for partial differential equations}.
\newblock \bibinfo{journal}{\emph{Proceedings of the National Academy of Sciences}} \bibinfo{volume}{116}, \bibinfo{number}{31} (\bibinfo{year}{2019}), \bibinfo{pages}{15344--15349}.
\newblock


\bibitem[Bastek et~al\mbox{.}(2024)]%
        {bastek2024physics}
\bibfield{author}{\bibinfo{person}{Jan-Hendrik Bastek}, \bibinfo{person}{WaiChing Sun}, {and} \bibinfo{person}{Dennis~M Kochmann}.} \bibinfo{year}{2024}\natexlab{}.
\newblock \showarticletitle{Physics-Informed Diffusion Models}.
\newblock \bibinfo{journal}{\emph{arXiv preprint arXiv:2403.14404}} (\bibinfo{year}{2024}).
\newblock


\bibitem[Bevacqua et~al\mbox{.}(2023)]%
        {bevacqua2023advancing}
\bibfield{author}{\bibinfo{person}{Emanuele Bevacqua}, \bibinfo{person}{Laura Suarez-Gutierrez}, \bibinfo{person}{Agla{\'e} J{\'e}z{\'e}quel}, \bibinfo{person}{Flavio Lehner}, \bibinfo{person}{Mathieu Vrac}, \bibinfo{person}{Pascal Yiou}, {and} \bibinfo{person}{Jakob Zscheischler}.} \bibinfo{year}{2023}\natexlab{}.
\newblock \showarticletitle{Advancing research on compound weather and climate events via large ensemble model simulations}.
\newblock \bibinfo{journal}{\emph{Nature Communications}} \bibinfo{volume}{14}, \bibinfo{number}{1} (\bibinfo{year}{2023}), \bibinfo{pages}{2145}.
\newblock


\bibitem[Bhatia et~al\mbox{.}(2021)]%
        {bhatia2021machine}
\bibfield{author}{\bibinfo{person}{Harsh Bhatia}, \bibinfo{person}{Timothy~S Carpenter}, \bibinfo{person}{Helgi~I Ing{\'o}lfsson}, \bibinfo{person}{Gautham Dharuman}, \bibinfo{person}{Piyush Karande}, \bibinfo{person}{Shusen Liu}, \bibinfo{person}{Tomas Oppelstrup}, \bibinfo{person}{Chris Neale}, \bibinfo{person}{Felice~C Lightstone}, \bibinfo{person}{Brian Van~Essen}, {et~al\mbox{.}}} \bibinfo{year}{2021}\natexlab{}.
\newblock \showarticletitle{Machine-learning-based dynamic-importance sampling for adaptive multiscale simulations}.
\newblock \bibinfo{journal}{\emph{Nature Machine Intelligence}} \bibinfo{volume}{3}, \bibinfo{number}{5} (\bibinfo{year}{2021}), \bibinfo{pages}{401--409}.
\newblock


\bibitem[Brooks et~al\mbox{.}(2009)]%
        {brooks2009charmm}
\bibfield{author}{\bibinfo{person}{Bernard~R Brooks}, \bibinfo{person}{Charles~L Brooks~III}, \bibinfo{person}{Alexander~D Mackerell~Jr}, \bibinfo{person}{Lennart Nilsson}, \bibinfo{person}{Robert~J Petrella}, \bibinfo{person}{Beno{\^\i}t Roux}, \bibinfo{person}{Youngdo Won}, \bibinfo{person}{Georgios Archontis}, \bibinfo{person}{Christian Bartels}, \bibinfo{person}{Stefan Boresch}, {et~al\mbox{.}}} \bibinfo{year}{2009}\natexlab{}.
\newblock \showarticletitle{CHARMM: the biomolecular simulation program}.
\newblock \bibinfo{journal}{\emph{Journal of computational chemistry}} \bibinfo{volume}{30}, \bibinfo{number}{10} (\bibinfo{year}{2009}), \bibinfo{pages}{1545--1614}.
\newblock


\bibitem[Brown et~al\mbox{.}(2015)]%
        {brown2015future}
\bibfield{author}{\bibinfo{person}{Casey~M Brown}, \bibinfo{person}{Jay~R Lund}, \bibinfo{person}{Ximing Cai}, \bibinfo{person}{Patrick~M Reed}, \bibinfo{person}{Edith~A Zagona}, \bibinfo{person}{Avi Ostfeld}, \bibinfo{person}{Jim Hall}, \bibinfo{person}{Gregory~W Characklis}, \bibinfo{person}{Winston Yu}, {and} \bibinfo{person}{Levi Brekke}.} \bibinfo{year}{2015}\natexlab{}.
\newblock \showarticletitle{The future of water resources systems analysis: Toward a scientific framework for sustainable water management}.
\newblock \bibinfo{journal}{\emph{Water resources research}} \bibinfo{volume}{51}, \bibinfo{number}{8} (\bibinfo{year}{2015}), \bibinfo{pages}{6110--6124}.
\newblock


\bibitem[Cao et~al\mbox{.}(2024)]%
        {cao2024survey}
\bibfield{author}{\bibinfo{person}{Hanqun Cao}, \bibinfo{person}{Cheng Tan}, \bibinfo{person}{Zhangyang Gao}, \bibinfo{person}{Yilun Xu}, \bibinfo{person}{Guangyong Chen}, \bibinfo{person}{Pheng-Ann Heng}, {and} \bibinfo{person}{Stan~Z Li}.} \bibinfo{year}{2024}\natexlab{}.
\newblock \showarticletitle{A survey on generative diffusion models}.
\newblock \bibinfo{journal}{\emph{IEEE Transactions on Knowledge and Data Engineering}} (\bibinfo{year}{2024}).
\newblock


\bibitem[Cash et~al\mbox{.}(2006)]%
        {cash2006scale}
\bibfield{author}{\bibinfo{person}{David~W Cash}, \bibinfo{person}{W~Neil Adger}, \bibinfo{person}{Fikret Berkes}, \bibinfo{person}{Po Garden}, \bibinfo{person}{Louis Lebel}, \bibinfo{person}{Per Olsson}, \bibinfo{person}{Lowell Pritchard}, {and} \bibinfo{person}{Oran Young}.} \bibinfo{year}{2006}\natexlab{}.
\newblock \showarticletitle{Scale and cross-scale dynamics: governance and information in a multilevel world}.
\newblock \bibinfo{journal}{\emph{Ecology and society}} \bibinfo{volume}{11}, \bibinfo{number}{2} (\bibinfo{year}{2006}).
\newblock


\bibitem[Champion et~al\mbox{.}(2019)]%
        {champion2019data}
\bibfield{author}{\bibinfo{person}{Kathleen Champion}, \bibinfo{person}{Bethany Lusch}, \bibinfo{person}{J~Nathan Kutz}, {and} \bibinfo{person}{Steven~L Brunton}.} \bibinfo{year}{2019}\natexlab{}.
\newblock \showarticletitle{Data-driven discovery of coordinates and governing equations}.
\newblock \bibinfo{journal}{\emph{Proceedings of the National Academy of Sciences}} \bibinfo{volume}{116}, \bibinfo{number}{45} (\bibinfo{year}{2019}), \bibinfo{pages}{22445--22451}.
\newblock


\bibitem[Chen et~al\mbox{.}(2018)]%
        {chen2018neural}
\bibfield{author}{\bibinfo{person}{Ricky~TQ Chen}, \bibinfo{person}{Yulia Rubanova}, \bibinfo{person}{Jesse Bettencourt}, {and} \bibinfo{person}{David~K Duvenaud}.} \bibinfo{year}{2018}\natexlab{}.
\newblock \showarticletitle{Neural ordinary differential equations}.
\newblock \bibinfo{journal}{\emph{Advances in neural information processing systems}}  \bibinfo{volume}{31} (\bibinfo{year}{2018}).
\newblock


\bibitem[Croitoru et~al\mbox{.}(2023)]%
        {croitoru2023diffusion}
\bibfield{author}{\bibinfo{person}{Florinel-Alin Croitoru}, \bibinfo{person}{Vlad Hondru}, \bibinfo{person}{Radu~Tudor Ionescu}, {and} \bibinfo{person}{Mubarak Shah}.} \bibinfo{year}{2023}\natexlab{}.
\newblock \showarticletitle{Diffusion models in vision: A survey}.
\newblock \bibinfo{journal}{\emph{IEEE Transactions on Pattern Analysis and Machine Intelligence}} \bibinfo{volume}{45}, \bibinfo{number}{9} (\bibinfo{year}{2023}), \bibinfo{pages}{10850--10869}.
\newblock


\bibitem[Dhariwal and Nichol(2021)]%
        {dhariwal2021diffusion}
\bibfield{author}{\bibinfo{person}{Prafulla Dhariwal} {and} \bibinfo{person}{Alexander Nichol}.} \bibinfo{year}{2021}\natexlab{}.
\newblock \showarticletitle{Diffusion models beat gans on image synthesis}.
\newblock \bibinfo{journal}{\emph{Advances in neural information processing systems}}  \bibinfo{volume}{34} (\bibinfo{year}{2021}), \bibinfo{pages}{8780--8794}.
\newblock


\bibitem[Ding et~al\mbox{.}(2025)]%
        {ding2025artificial}
\bibfield{author}{\bibinfo{person}{Jingtao Ding}, \bibinfo{person}{Yu Zheng}, \bibinfo{person}{Huandong Wang}, \bibinfo{person}{Carlo~Vittorio Cannistraci}, \bibinfo{person}{Jianxi Gao}, \bibinfo{person}{Yong Li}, {and} \bibinfo{person}{Chuan Shi}.} \bibinfo{year}{2025}\natexlab{}.
\newblock \showarticletitle{Artificial intelligence for complex network: Potential, methodology and application}. In \bibinfo{booktitle}{\emph{Companion Proceedings of the ACM on Web Conference 2025}}. \bibinfo{pages}{5--8}.
\newblock


\bibitem[Fan et~al\mbox{.}(2024)]%
        {fan2024mg}
\bibfield{author}{\bibinfo{person}{Xinyao Fan}, \bibinfo{person}{Yueying Wu}, \bibinfo{person}{Chang Xu}, \bibinfo{person}{Yuhao Huang}, \bibinfo{person}{Weiqing Liu}, {and} \bibinfo{person}{Jiang Bian}.} \bibinfo{year}{2024}\natexlab{}.
\newblock \showarticletitle{MG-TSD: Multi-granularity time series diffusion models with guided learning process}.
\newblock \bibinfo{journal}{\emph{arXiv preprint arXiv:2403.05751}} (\bibinfo{year}{2024}).
\newblock


\bibitem[Federici et~al\mbox{.}(2023)]%
        {federici2023latent}
\bibfield{author}{\bibinfo{person}{Marco Federici}, \bibinfo{person}{Patrick Forr{\'e}}, \bibinfo{person}{Ryota Tomioka}, {and} \bibinfo{person}{Bastiaan~S Veeling}.} \bibinfo{year}{2023}\natexlab{}.
\newblock \showarticletitle{Latent representation and simulation of markov processes via time-lagged information bottleneck}.
\newblock \bibinfo{journal}{\emph{arXiv preprint arXiv:2309.07200}} (\bibinfo{year}{2023}).
\newblock


\bibitem[Gao et~al\mbox{.}(2024)]%
        {gao2024generative}
\bibfield{author}{\bibinfo{person}{Han Gao}, \bibinfo{person}{Sebastian Kaltenbach}, {and} \bibinfo{person}{Petros Koumoutsakos}.} \bibinfo{year}{2024}\natexlab{}.
\newblock \showarticletitle{Generative learning for forecasting the dynamics of high-dimensional complex systems}.
\newblock \bibinfo{journal}{\emph{Nature Communications}} \bibinfo{volume}{15}, \bibinfo{number}{1} (\bibinfo{year}{2024}), \bibinfo{pages}{8904}.
\newblock


\bibitem[Gao(2024)]%
        {gao2024intrinsic}
\bibfield{author}{\bibinfo{person}{Jianxi Gao}.} \bibinfo{year}{2024}\natexlab{}.
\newblock \showarticletitle{Intrinsic simplicity of complex systems}.
\newblock \bibinfo{journal}{\emph{Nature Physics}} \bibinfo{volume}{20}, \bibinfo{number}{2} (\bibinfo{year}{2024}), \bibinfo{pages}{184--185}.
\newblock


\bibitem[Garc{\'\i}a-P{\'e}rez et~al\mbox{.}(2018)]%
        {garcia2018multiscale}
\bibfield{author}{\bibinfo{person}{Guillermo Garc{\'\i}a-P{\'e}rez}, \bibinfo{person}{Mari{\'a}n Bogu{\~n}{\'a}}, {and} \bibinfo{person}{M~{\'A}ngeles Serrano}.} \bibinfo{year}{2018}\natexlab{}.
\newblock \showarticletitle{Multiscale unfolding of real networks by geometric renormalization}.
\newblock \bibinfo{journal}{\emph{Nature Physics}} \bibinfo{volume}{14}, \bibinfo{number}{6} (\bibinfo{year}{2018}), \bibinfo{pages}{583--589}.
\newblock


\bibitem[Gupta et~al\mbox{.}(2022)]%
        {gupta2022non}
\bibfield{author}{\bibinfo{person}{Gaurav Gupta}, \bibinfo{person}{Xiongye Xiao}, \bibinfo{person}{Radu Balan}, {and} \bibinfo{person}{Paul Bogdan}.} \bibinfo{year}{2022}\natexlab{}.
\newblock \showarticletitle{Non-linear operator approximations for initial value problems}. In \bibinfo{booktitle}{\emph{International Conference on Learning Representations (ICLR)}}.
\newblock


\bibitem[Gupta et~al\mbox{.}(2021)]%
        {gupta2021multiwavelet}
\bibfield{author}{\bibinfo{person}{Gaurav Gupta}, \bibinfo{person}{Xiongye Xiao}, {and} \bibinfo{person}{Paul Bogdan}.} \bibinfo{year}{2021}\natexlab{}.
\newblock \showarticletitle{Multiwavelet-based operator learning for differential equations}.
\newblock \bibinfo{journal}{\emph{Advances in neural information processing systems}}  \bibinfo{volume}{34} (\bibinfo{year}{2021}), \bibinfo{pages}{24048--24062}.
\newblock


\bibitem[Haken(2006)]%
        {haken2006information}
\bibfield{author}{\bibinfo{person}{H Haken}.} \bibinfo{year}{2006}\natexlab{}.
\newblock \bibinfo{booktitle}{\emph{Information and self-organization: A macroscopic approach to complex systems}}.
\newblock \bibinfo{publisher}{Springer}.
\newblock


\bibitem[Haken(2012)]%
        {haken2012advanced}
\bibfield{author}{\bibinfo{person}{H Haken}.} \bibinfo{year}{2012}\natexlab{}.
\newblock \bibinfo{booktitle}{\emph{Advanced synergetics: instability hierarchies of self-organizing systems and devices}}.
\newblock \bibinfo{publisher}{Springer}.
\newblock


\bibitem[Ho et~al\mbox{.}(2020)]%
        {ho2020denoising}
\bibfield{author}{\bibinfo{person}{Jonathan Ho}, \bibinfo{person}{Ajay Jain}, {and} \bibinfo{person}{Pieter Abbeel}.} \bibinfo{year}{2020}\natexlab{}.
\newblock \showarticletitle{Denoising diffusion probabilistic models}.
\newblock \bibinfo{journal}{\emph{Advances in neural information processing systems}}  \bibinfo{volume}{33} (\bibinfo{year}{2020}), \bibinfo{pages}{6840--6851}.
\newblock


\bibitem[Hore and Ziou(2010)]%
        {hore2010image}
\bibfield{author}{\bibinfo{person}{Alain Hore} {and} \bibinfo{person}{Djemel Ziou}.} \bibinfo{year}{2010}\natexlab{}.
\newblock \showarticletitle{Image quality metrics: PSNR vs. SSIM}. In \bibinfo{booktitle}{\emph{2010 20th international conference on pattern recognition}}. IEEE, \bibinfo{pages}{2366--2369}.
\newblock


\bibitem[Jin et~al\mbox{.}(2024)]%
        {jin2024pyramidal}
\bibfield{author}{\bibinfo{person}{Yang Jin}, \bibinfo{person}{Zhicheng Sun}, \bibinfo{person}{Ningyuan Li}, \bibinfo{person}{Kun Xu}, \bibinfo{person}{Hao Jiang}, \bibinfo{person}{Nan Zhuang}, \bibinfo{person}{Quzhe Huang}, \bibinfo{person}{Yang Song}, \bibinfo{person}{Yadong Mu}, {and} \bibinfo{person}{Zhouchen Lin}.} \bibinfo{year}{2024}\natexlab{}.
\newblock \showarticletitle{Pyramidal flow matching for efficient video generative modeling}.
\newblock \bibinfo{journal}{\emph{arXiv preprint arXiv:2410.05954}} (\bibinfo{year}{2024}).
\newblock


\bibitem[Karras et~al\mbox{.}(2022)]%
        {karras2022elucidating}
\bibfield{author}{\bibinfo{person}{Tero Karras}, \bibinfo{person}{Miika Aittala}, \bibinfo{person}{Timo Aila}, {and} \bibinfo{person}{Samuli Laine}.} \bibinfo{year}{2022}\natexlab{}.
\newblock \showarticletitle{Elucidating the design space of diffusion-based generative models}.
\newblock \bibinfo{journal}{\emph{Advances in neural information processing systems}}  \bibinfo{volume}{35} (\bibinfo{year}{2022}), \bibinfo{pages}{26565--26577}.
\newblock


\bibitem[Koch-Janusz and Ringel(2018)]%
        {koch2018mutual}
\bibfield{author}{\bibinfo{person}{Maciej Koch-Janusz} {and} \bibinfo{person}{Zohar Ringel}.} \bibinfo{year}{2018}\natexlab{}.
\newblock \showarticletitle{Mutual information, neural networks and the renormalization group}.
\newblock \bibinfo{journal}{\emph{Nature Physics}} \bibinfo{volume}{14}, \bibinfo{number}{6} (\bibinfo{year}{2018}), \bibinfo{pages}{578--582}.
\newblock


\bibitem[Kochkov et~al\mbox{.}(2021)]%
        {kochkov2021machine}
\bibfield{author}{\bibinfo{person}{Dmitrii Kochkov}, \bibinfo{person}{Jamie~A Smith}, \bibinfo{person}{Ayya Alieva}, \bibinfo{person}{Qing Wang}, \bibinfo{person}{Michael~P Brenner}, {and} \bibinfo{person}{Stephan Hoyer}.} \bibinfo{year}{2021}\natexlab{}.
\newblock \showarticletitle{Machine learning--accelerated computational fluid dynamics}.
\newblock \bibinfo{journal}{\emph{Proceedings of the National Academy of Sciences}} \bibinfo{volume}{118}, \bibinfo{number}{21} (\bibinfo{year}{2021}), \bibinfo{pages}{e2101784118}.
\newblock


\bibitem[Kontolati et~al\mbox{.}(2024)]%
        {kontolati2024learning}
\bibfield{author}{\bibinfo{person}{Katiana Kontolati}, \bibinfo{person}{Somdatta Goswami}, \bibinfo{person}{George Em~Karniadakis}, {and} \bibinfo{person}{Michael~D Shields}.} \bibinfo{year}{2024}\natexlab{}.
\newblock \showarticletitle{Learning nonlinear operators in latent spaces for real-time predictions of complex dynamics in physical systems}.
\newblock \bibinfo{journal}{\emph{Nature Communications}} \bibinfo{volume}{15}, \bibinfo{number}{1} (\bibinfo{year}{2024}), \bibinfo{pages}{5101}.
\newblock


\bibitem[Lee et~al\mbox{.}(2022)]%
        {lee2022autoregressive}
\bibfield{author}{\bibinfo{person}{Doyup Lee}, \bibinfo{person}{Chiheon Kim}, \bibinfo{person}{Saehoon Kim}, \bibinfo{person}{Minsu Cho}, {and} \bibinfo{person}{Wook-Shin Han}.} \bibinfo{year}{2022}\natexlab{}.
\newblock \showarticletitle{Autoregressive image generation using residual quantization}. In \bibinfo{booktitle}{\emph{Proceedings of the IEEE/CVF Conference on Computer Vision and Pattern Recognition}}. \bibinfo{pages}{11523--11532}.
\newblock


\bibitem[Lee et~al\mbox{.}(2020)]%
        {lee2020coarse}
\bibfield{author}{\bibinfo{person}{Seungjoon Lee}, \bibinfo{person}{Mahdi Kooshkbaghi}, \bibinfo{person}{Konstantinos Spiliotis}, \bibinfo{person}{Constantinos~I Siettos}, {and} \bibinfo{person}{Ioannis~G Kevrekidis}.} \bibinfo{year}{2020}\natexlab{}.
\newblock \showarticletitle{Coarse-scale PDEs from fine-scale observations via machine learning}.
\newblock \bibinfo{journal}{\emph{Chaos: An Interdisciplinary Journal of Nonlinear Science}} \bibinfo{volume}{30}, \bibinfo{number}{1} (\bibinfo{year}{2020}).
\newblock


\bibitem[Li et~al\mbox{.}(2004)]%
        {li2004multi}
\bibfield{author}{\bibinfo{person}{Jinghai Li}, \bibinfo{person}{Jiayuan Zhang}, \bibinfo{person}{Wei Ge}, {and} \bibinfo{person}{Xinhua Liu}.} \bibinfo{year}{2004}\natexlab{}.
\newblock \showarticletitle{Multi-scale methodology for complex systems}.
\newblock \bibinfo{journal}{\emph{Chemical engineering science}} \bibinfo{volume}{59}, \bibinfo{number}{8-9} (\bibinfo{year}{2004}), \bibinfo{pages}{1687--1700}.
\newblock


\bibitem[Li et~al\mbox{.}(2025a)]%
        {li2025predicting2}
\bibfield{author}{\bibinfo{person}{Ruikun Li}, \bibinfo{person}{Huandong Wang}, \bibinfo{person}{Jingtao Ding}, \bibinfo{person}{Yuan Yuan}, \bibinfo{person}{Qingmin Liao}, {and} \bibinfo{person}{Yong Li}.} \bibinfo{year}{2025}\natexlab{a}.
\newblock \showarticletitle{Predicting Dynamical Systems across Environments via Diffusive Model Weight Generation}.
\newblock \bibinfo{journal}{\emph{arXiv preprint arXiv:2505.13919}} (\bibinfo{year}{2025}).
\newblock


\bibitem[Li et~al\mbox{.}(2023)]%
        {li2023learning}
\bibfield{author}{\bibinfo{person}{Ruikun Li}, \bibinfo{person}{Huandong Wang}, {and} \bibinfo{person}{Yong Li}.} \bibinfo{year}{2023}\natexlab{}.
\newblock \showarticletitle{Learning slow and fast system dynamics via automatic separation of time scales}. In \bibinfo{booktitle}{\emph{Proceedings of the 29th ACM SIGKDD Conference on Knowledge Discovery and Data Mining}}. \bibinfo{pages}{4380--4390}.
\newblock


\bibitem[Li et~al\mbox{.}(2025b)]%
        {li2025predicting}
\bibfield{author}{\bibinfo{person}{Ruikun Li}, \bibinfo{person}{Huandong Wang}, \bibinfo{person}{Qingmin Liao}, {and} \bibinfo{person}{Yong Li}.} \bibinfo{year}{2025}\natexlab{b}.
\newblock \showarticletitle{Predicting the Energy Landscape of Stochastic Dynamical System via Physics-informed Self-supervised Learning}.
\newblock \bibinfo{journal}{\emph{arXiv preprint arXiv:2502.16828}} (\bibinfo{year}{2025}).
\newblock


\bibitem[Li et~al\mbox{.}(2024a)]%
        {li2024synthetic}
\bibfield{author}{\bibinfo{person}{Tianyi Li}, \bibinfo{person}{Luca Biferale}, \bibinfo{person}{Fabio Bonaccorso}, \bibinfo{person}{Martino~Andrea Scarpolini}, {and} \bibinfo{person}{Michele Buzzicotti}.} \bibinfo{year}{2024}\natexlab{a}.
\newblock \showarticletitle{Synthetic Lagrangian turbulence by generative diffusion models}.
\newblock \bibinfo{journal}{\emph{Nature Machine Intelligence}} (\bibinfo{year}{2024}), \bibinfo{pages}{1--11}.
\newblock


\bibitem[Li et~al\mbox{.}(2024b)]%
        {li2024learning}
\bibfield{author}{\bibinfo{person}{Zeyu Li}, \bibinfo{person}{Wang Han}, \bibinfo{person}{Yue Zhang}, \bibinfo{person}{Qingfei Fu}, \bibinfo{person}{Jingxuan Li}, \bibinfo{person}{Lizi Qin}, \bibinfo{person}{Ruoyu Dong}, \bibinfo{person}{Hao Sun}, \bibinfo{person}{Yue Deng}, {and} \bibinfo{person}{Lijun Yang}.} \bibinfo{year}{2024}\natexlab{b}.
\newblock \showarticletitle{Learning spatiotemporal dynamics with a pretrained generative model}.
\newblock \bibinfo{journal}{\emph{Nature Machine Intelligence}} \bibinfo{volume}{6}, \bibinfo{number}{12} (\bibinfo{year}{2024}), \bibinfo{pages}{1566--1579}.
\newblock


\bibitem[Li et~al\mbox{.}({[n.\,d.]})]%
        {lifourier}
\bibfield{author}{\bibinfo{person}{Zongyi Li}, \bibinfo{person}{Nikola~Borislavov Kovachki}, \bibinfo{person}{Kamyar Azizzadenesheli}, \bibinfo{person}{Kaushik Bhattacharya}, \bibinfo{person}{Andrew Stuart}, \bibinfo{person}{Anima Anandkumar}, {et~al\mbox{.}}} \bibinfo{year}{[n.\,d.]}\natexlab{}.
\newblock \showarticletitle{Fourier Neural Operator for Parametric Partial Differential Equations}. In \bibinfo{booktitle}{\emph{International Conference on Learning Representations}}.
\newblock


\bibitem[Lienen et~al\mbox{.}(2023)]%
        {lienen2023zero}
\bibfield{author}{\bibinfo{person}{Marten Lienen}, \bibinfo{person}{David L{\"u}dke}, \bibinfo{person}{Jan Hansen-Palmus}, {and} \bibinfo{person}{Stephan G{\"u}nnemann}.} \bibinfo{year}{2023}\natexlab{}.
\newblock \showarticletitle{From zero to turbulence: Generative modeling for 3d flow simulation}.
\newblock \bibinfo{journal}{\emph{arXiv preprint arXiv:2306.01776}} (\bibinfo{year}{2023}).
\newblock


\bibitem[Lino et~al\mbox{.}(2021)]%
        {lino2021simulating}
\bibfield{author}{\bibinfo{person}{Mario Lino}, \bibinfo{person}{Chris Cantwell}, \bibinfo{person}{Anil~A Bharath}, {and} \bibinfo{person}{Stathi Fotiadis}.} \bibinfo{year}{2021}\natexlab{}.
\newblock \showarticletitle{Simulating continuum mechanics with multi-scale graph neural networks}.
\newblock \bibinfo{journal}{\emph{arXiv preprint arXiv:2106.04900}} (\bibinfo{year}{2021}).
\newblock


\bibitem[Lu et~al\mbox{.}(2021)]%
        {lu2021learning}
\bibfield{author}{\bibinfo{person}{Lu Lu}, \bibinfo{person}{Pengzhan Jin}, \bibinfo{person}{Guofei Pang}, \bibinfo{person}{Zhongqiang Zhang}, {and} \bibinfo{person}{George~Em Karniadakis}.} \bibinfo{year}{2021}\natexlab{}.
\newblock \showarticletitle{Learning nonlinear operators via DeepONet based on the universal approximation theorem of operators}.
\newblock \bibinfo{journal}{\emph{Nature machine intelligence}} \bibinfo{volume}{3}, \bibinfo{number}{3} (\bibinfo{year}{2021}), \bibinfo{pages}{218--229}.
\newblock


\bibitem[Lusch et~al\mbox{.}(2018)]%
        {lusch2018deep}
\bibfield{author}{\bibinfo{person}{Bethany Lusch}, \bibinfo{person}{J~Nathan Kutz}, {and} \bibinfo{person}{Steven~L Brunton}.} \bibinfo{year}{2018}\natexlab{}.
\newblock \showarticletitle{Deep learning for universal linear embeddings of nonlinear dynamics}.
\newblock \bibinfo{journal}{\emph{Nature communications}} \bibinfo{volume}{9}, \bibinfo{number}{1} (\bibinfo{year}{2018}), \bibinfo{pages}{4950}.
\newblock


\bibitem[Mardt et~al\mbox{.}(2018)]%
        {mardt2018vampnets}
\bibfield{author}{\bibinfo{person}{Andreas Mardt}, \bibinfo{person}{Luca Pasquali}, \bibinfo{person}{Hao Wu}, {and} \bibinfo{person}{Frank No{\'e}}.} \bibinfo{year}{2018}\natexlab{}.
\newblock \showarticletitle{VAMPnets for deep learning of molecular kinetics}.
\newblock \bibinfo{journal}{\emph{Nature communications}} \bibinfo{volume}{9}, \bibinfo{number}{1} (\bibinfo{year}{2018}), \bibinfo{pages}{5}.
\newblock


\bibitem[Ohana et~al\mbox{.}(2024)]%
        {ohana2024well}
\bibfield{author}{\bibinfo{person}{Ruben Ohana}, \bibinfo{person}{Michael McCabe}, \bibinfo{person}{Lucas Meyer}, \bibinfo{person}{Rudy Morel}, \bibinfo{person}{Fruzsina~J Agocs}, \bibinfo{person}{Miguel Beneitez}, \bibinfo{person}{Marsha Berger}, \bibinfo{person}{Blakesley Burkhart}, \bibinfo{person}{Stuart~B Dalziel}, \bibinfo{person}{Drummond~B Fielding}, {et~al\mbox{.}}} \bibinfo{year}{2024}\natexlab{}.
\newblock \showarticletitle{The Well: a Large-Scale Collection of Diverse Physics Simulations for Machine Learning}.
\newblock \bibinfo{journal}{\emph{arXiv preprint arXiv:2412.00568}} (\bibinfo{year}{2024}).
\newblock


\bibitem[Peng et~al\mbox{.}(2021)]%
        {peng2021multiscale}
\bibfield{author}{\bibinfo{person}{Grace~CY Peng}, \bibinfo{person}{Mark Alber}, \bibinfo{person}{Adrian Buganza~Tepole}, \bibinfo{person}{William~R Cannon}, \bibinfo{person}{Suvranu De}, \bibinfo{person}{Savador Dura-Bernal}, \bibinfo{person}{Krishna Garikipati}, \bibinfo{person}{George Karniadakis}, \bibinfo{person}{William~W Lytton}, \bibinfo{person}{Paris Perdikaris}, {et~al\mbox{.}}} \bibinfo{year}{2021}\natexlab{}.
\newblock \showarticletitle{Multiscale modeling meets machine learning: What can we learn?}
\newblock \bibinfo{journal}{\emph{Archives of Computational Methods in Engineering}}  \bibinfo{volume}{28} (\bibinfo{year}{2021}), \bibinfo{pages}{1017--1037}.
\newblock


\bibitem[Preechakul et~al\mbox{.}(2022)]%
        {preechakul2022diffusion}
\bibfield{author}{\bibinfo{person}{Konpat Preechakul}, \bibinfo{person}{Nattanat Chatthee}, \bibinfo{person}{Suttisak Wizadwongsa}, {and} \bibinfo{person}{Supasorn Suwajanakorn}.} \bibinfo{year}{2022}\natexlab{}.
\newblock \showarticletitle{Diffusion autoencoders: Toward a meaningful and decodable representation}. In \bibinfo{booktitle}{\emph{Proceedings of the IEEE/CVF conference on computer vision and pattern recognition}}. \bibinfo{pages}{10619--10629}.
\newblock


\bibitem[Preisler and Westerling(2007)]%
        {preisler2007statistical}
\bibfield{author}{\bibinfo{person}{Haiganoush~K Preisler} {and} \bibinfo{person}{Anthony~L Westerling}.} \bibinfo{year}{2007}\natexlab{}.
\newblock \showarticletitle{Statistical model for forecasting monthly large wildfire events in western United States}.
\newblock \bibinfo{journal}{\emph{Journal of Applied Meteorology and Climatology}} \bibinfo{volume}{46}, \bibinfo{number}{7} (\bibinfo{year}{2007}), \bibinfo{pages}{1020--1030}.
\newblock


\bibitem[Prigogine and Nicolis(1967)]%
        {prigogine1967symmetry}
\bibfield{author}{\bibinfo{person}{Ilya Prigogine} {and} \bibinfo{person}{Gr{\'e}goire Nicolis}.} \bibinfo{year}{1967}\natexlab{}.
\newblock \showarticletitle{On symmetry-breaking instabilities in dissipative systems}.
\newblock \bibinfo{journal}{\emph{The Journal of Chemical Physics}} \bibinfo{volume}{46}, \bibinfo{number}{9} (\bibinfo{year}{1967}), \bibinfo{pages}{3542--3550}.
\newblock


\bibitem[Rao et~al\mbox{.}(2023)]%
        {rao2023encoding}
\bibfield{author}{\bibinfo{person}{Chengping Rao}, \bibinfo{person}{Pu Ren}, \bibinfo{person}{Qi Wang}, \bibinfo{person}{Oral Buyukozturk}, \bibinfo{person}{Hao Sun}, {and} \bibinfo{person}{Yang Liu}.} \bibinfo{year}{2023}\natexlab{}.
\newblock \showarticletitle{Encoding physics to learn reaction--diffusion processes}.
\newblock \bibinfo{journal}{\emph{Nature Machine Intelligence}} \bibinfo{volume}{5}, \bibinfo{number}{7} (\bibinfo{year}{2023}), \bibinfo{pages}{765--779}.
\newblock


\bibitem[Rombach et~al\mbox{.}(2022)]%
        {rombach2022high}
\bibfield{author}{\bibinfo{person}{Robin Rombach}, \bibinfo{person}{Andreas Blattmann}, \bibinfo{person}{Dominik Lorenz}, \bibinfo{person}{Patrick Esser}, {and} \bibinfo{person}{Bj{\"o}rn Ommer}.} \bibinfo{year}{2022}\natexlab{}.
\newblock \showarticletitle{High-resolution image synthesis with latent diffusion models}. In \bibinfo{booktitle}{\emph{Proceedings of the IEEE/CVF conference on computer vision and pattern recognition}}. \bibinfo{pages}{10684--10695}.
\newblock


\bibitem[Ronneberger et~al\mbox{.}(2015)]%
        {ronneberger2015u}
\bibfield{author}{\bibinfo{person}{Olaf Ronneberger}, \bibinfo{person}{Philipp Fischer}, {and} \bibinfo{person}{Thomas Brox}.} \bibinfo{year}{2015}\natexlab{}.
\newblock \showarticletitle{U-net: Convolutional networks for biomedical image segmentation}. In \bibinfo{booktitle}{\emph{Medical image computing and computer-assisted intervention--MICCAI 2015: 18th international conference, Munich, Germany, October 5-9, 2015, proceedings, part III 18}}. Springer, \bibinfo{pages}{234--241}.
\newblock


\bibitem[R{\"u}hling~Cachay et~al\mbox{.}(2024)]%
        {ruhling2024dyffusion}
\bibfield{author}{\bibinfo{person}{Salva R{\"u}hling~Cachay}, \bibinfo{person}{Bo Zhao}, \bibinfo{person}{Hailey Joren}, {and} \bibinfo{person}{Rose Yu}.} \bibinfo{year}{2024}\natexlab{}.
\newblock \showarticletitle{Dyffusion: A dynamics-informed diffusion model for spatiotemporal forecasting}.
\newblock \bibinfo{journal}{\emph{Advances in Neural Information Processing Systems}}  \bibinfo{volume}{36} (\bibinfo{year}{2024}).
\newblock


\bibitem[Seifner and S{\'a}nchez(2023)]%
        {seifner2023neural}
\bibfield{author}{\bibinfo{person}{Patrick Seifner} {and} \bibinfo{person}{Rams{\'e}s~J S{\'a}nchez}.} \bibinfo{year}{2023}\natexlab{}.
\newblock \showarticletitle{Neural Markov jump processes}. In \bibinfo{booktitle}{\emph{International Conference on Machine Learning}}. PMLR, \bibinfo{pages}{30523--30552}.
\newblock


\bibitem[Shen et~al\mbox{.}(2024)]%
        {shen2024multi}
\bibfield{author}{\bibinfo{person}{Lifeng Shen}, \bibinfo{person}{Weiyu Chen}, {and} \bibinfo{person}{James Kwok}.} \bibinfo{year}{2024}\natexlab{}.
\newblock \showarticletitle{Multi-Resolution Diffusion Models for Time Series Forecasting}. In \bibinfo{booktitle}{\emph{The Twelfth International Conference on Learning Representations}}.
\newblock


\bibitem[Sheng et~al\mbox{.}(2025)]%
        {sheng2025collaborative}
\bibfield{author}{\bibinfo{person}{Zhi Sheng}, \bibinfo{person}{Yuan Yuan}, \bibinfo{person}{Yudi Zhang}, \bibinfo{person}{Depeng Jin}, {and} \bibinfo{person}{Yong Li}.} \bibinfo{year}{2025}\natexlab{}.
\newblock \showarticletitle{Collaborative Deterministic-Probabilistic Forecasting for Real-World Spatiotemporal Systems}.
\newblock \bibinfo{journal}{\emph{arXiv preprint arXiv:2502.11013}} (\bibinfo{year}{2025}).
\newblock


\bibitem[Shi et~al\mbox{.}(2015)]%
        {shi2015convolutional}
\bibfield{author}{\bibinfo{person}{Xingjian Shi}, \bibinfo{person}{Zhourong Chen}, \bibinfo{person}{Hao Wang}, \bibinfo{person}{Dit-Yan Yeung}, \bibinfo{person}{Wai-Kin Wong}, {and} \bibinfo{person}{Wang-chun Woo}.} \bibinfo{year}{2015}\natexlab{}.
\newblock \showarticletitle{Convolutional LSTM network: A machine learning approach for precipitation nowcasting}.
\newblock \bibinfo{journal}{\emph{Advances in neural information processing systems}}  \bibinfo{volume}{28} (\bibinfo{year}{2015}).
\newblock


\bibitem[Shu et~al\mbox{.}(2023)]%
        {shu2023physics}
\bibfield{author}{\bibinfo{person}{Dule Shu}, \bibinfo{person}{Zijie Li}, {and} \bibinfo{person}{Amir~Barati Farimani}.} \bibinfo{year}{2023}\natexlab{}.
\newblock \showarticletitle{A physics-informed diffusion model for high-fidelity flow field reconstruction}.
\newblock \bibinfo{journal}{\emph{J. Comput. Phys.}}  \bibinfo{volume}{478} (\bibinfo{year}{2023}), \bibinfo{pages}{111972}.
\newblock


\bibitem[Song et~al\mbox{.}(2020a)]%
        {song2020denoising}
\bibfield{author}{\bibinfo{person}{Jiaming Song}, \bibinfo{person}{Chenlin Meng}, {and} \bibinfo{person}{Stefano Ermon}.} \bibinfo{year}{2020}\natexlab{a}.
\newblock \showarticletitle{Denoising diffusion implicit models}.
\newblock \bibinfo{journal}{\emph{arXiv preprint arXiv:2010.02502}} (\bibinfo{year}{2020}).
\newblock


\bibitem[Song and Ermon(2019)]%
        {song2019generative}
\bibfield{author}{\bibinfo{person}{Yang Song} {and} \bibinfo{person}{Stefano Ermon}.} \bibinfo{year}{2019}\natexlab{}.
\newblock \showarticletitle{Generative modeling by estimating gradients of the data distribution}.
\newblock \bibinfo{journal}{\emph{Advances in neural information processing systems}}  \bibinfo{volume}{32} (\bibinfo{year}{2019}).
\newblock


\bibitem[Song et~al\mbox{.}(2020b)]%
        {song2020score}
\bibfield{author}{\bibinfo{person}{Yang Song}, \bibinfo{person}{Jascha Sohl-Dickstein}, \bibinfo{person}{Diederik~P Kingma}, \bibinfo{person}{Abhishek Kumar}, \bibinfo{person}{Stefano Ermon}, {and} \bibinfo{person}{Ben Poole}.} \bibinfo{year}{2020}\natexlab{b}.
\newblock \showarticletitle{Score-based generative modeling through stochastic differential equations}.
\newblock \bibinfo{journal}{\emph{arXiv preprint arXiv:2011.13456}} (\bibinfo{year}{2020}).
\newblock


\bibitem[Takamoto et~al\mbox{.}(2022)]%
        {takamoto2022pdebench}
\bibfield{author}{\bibinfo{person}{Makoto Takamoto}, \bibinfo{person}{Timothy Praditia}, \bibinfo{person}{Raphael Leiteritz}, \bibinfo{person}{Daniel MacKinlay}, \bibinfo{person}{Francesco Alesiani}, \bibinfo{person}{Dirk Pfl{\"u}ger}, {and} \bibinfo{person}{Mathias Niepert}.} \bibinfo{year}{2022}\natexlab{}.
\newblock \showarticletitle{Pdebench: An extensive benchmark for scientific machine learning}.
\newblock \bibinfo{journal}{\emph{Advances in Neural Information Processing Systems}}  \bibinfo{volume}{35} (\bibinfo{year}{2022}), \bibinfo{pages}{1596--1611}.
\newblock


\bibitem[Thibeault et~al\mbox{.}(2024)]%
        {thibeault2024low}
\bibfield{author}{\bibinfo{person}{Vincent Thibeault}, \bibinfo{person}{Antoine Allard}, {and} \bibinfo{person}{Patrick Desrosiers}.} \bibinfo{year}{2024}\natexlab{}.
\newblock \showarticletitle{The low-rank hypothesis of complex systems}.
\newblock \bibinfo{journal}{\emph{Nature Physics}} \bibinfo{volume}{20}, \bibinfo{number}{2} (\bibinfo{year}{2024}), \bibinfo{pages}{294--302}.
\newblock


\bibitem[Tian et~al\mbox{.}(2024)]%
        {tian2024visual}
\bibfield{author}{\bibinfo{person}{Keyu Tian}, \bibinfo{person}{Yi Jiang}, \bibinfo{person}{Zehuan Yuan}, \bibinfo{person}{Bingyue Peng}, {and} \bibinfo{person}{Liwei Wang}.} \bibinfo{year}{2024}\natexlab{}.
\newblock \showarticletitle{Visual autoregressive modeling: Scalable image generation via next-scale prediction}.
\newblock \bibinfo{journal}{\emph{arXiv preprint arXiv:2404.02905}} (\bibinfo{year}{2024}).
\newblock


\bibitem[Tumanyan et~al\mbox{.}(2023)]%
        {tumanyan2023plug}
\bibfield{author}{\bibinfo{person}{Narek Tumanyan}, \bibinfo{person}{Michal Geyer}, \bibinfo{person}{Shai Bagon}, {and} \bibinfo{person}{Tali Dekel}.} \bibinfo{year}{2023}\natexlab{}.
\newblock \showarticletitle{Plug-and-play diffusion features for text-driven image-to-image translation}. In \bibinfo{booktitle}{\emph{Proceedings of the IEEE/CVF Conference on Computer Vision and Pattern Recognition}}. \bibinfo{pages}{1921--1930}.
\newblock


\bibitem[Veli{\v{c}}kovi{\'c} et~al\mbox{.}(2017)]%
        {velivckovic2017graph}
\bibfield{author}{\bibinfo{person}{Petar Veli{\v{c}}kovi{\'c}}, \bibinfo{person}{Guillem Cucurull}, \bibinfo{person}{Arantxa Casanova}, \bibinfo{person}{Adriana Romero}, \bibinfo{person}{Pietro Lio}, {and} \bibinfo{person}{Yoshua Bengio}.} \bibinfo{year}{2017}\natexlab{}.
\newblock \showarticletitle{Graph attention networks}.
\newblock \bibinfo{journal}{\emph{arXiv preprint arXiv:1710.10903}} (\bibinfo{year}{2017}).
\newblock


\bibitem[Veličković et~al\mbox{.}(2018)]%
        {gat_ref}
\bibfield{author}{\bibinfo{person}{Petar Veličković}, \bibinfo{person}{Guillem Cucurull}, \bibinfo{person}{Arantxa Casanova}, \bibinfo{person}{Adriana Romero}, \bibinfo{person}{Pietro Liò}, {and} \bibinfo{person}{Yoshua Bengio}.} \bibinfo{year}{2018}\natexlab{}.
\newblock \showarticletitle{Graph attention networks}.
\newblock \bibinfo{journal}{\emph{ICLR}} (\bibinfo{year}{2018}).
\newblock


\bibitem[Villegas et~al\mbox{.}(2023)]%
        {villegas2023laplacian}
\bibfield{author}{\bibinfo{person}{Pablo Villegas}, \bibinfo{person}{Tommaso Gili}, \bibinfo{person}{Guido Caldarelli}, {and} \bibinfo{person}{Andrea Gabrielli}.} \bibinfo{year}{2023}\natexlab{}.
\newblock \showarticletitle{Laplacian renormalization group for heterogeneous networks}.
\newblock \bibinfo{journal}{\emph{Nature Physics}} \bibinfo{volume}{19}, \bibinfo{number}{3} (\bibinfo{year}{2023}), \bibinfo{pages}{445--450}.
\newblock


\bibitem[Vlachas et~al\mbox{.}(2022)]%
        {vlachas2022multiscale}
\bibfield{author}{\bibinfo{person}{Pantelis~R Vlachas}, \bibinfo{person}{Georgios Arampatzis}, \bibinfo{person}{Caroline Uhler}, {and} \bibinfo{person}{Petros Koumoutsakos}.} \bibinfo{year}{2022}\natexlab{}.
\newblock \showarticletitle{Multiscale simulations of complex systems by learning their effective dynamics}.
\newblock \bibinfo{journal}{\emph{Nature Machine Intelligence}} \bibinfo{volume}{4}, \bibinfo{number}{4} (\bibinfo{year}{2022}), \bibinfo{pages}{359--366}.
\newblock


\bibitem[Wang et~al\mbox{.}(2024b)]%
        {wang2024multi}
\bibfield{author}{\bibinfo{person}{Huandong Wang}, \bibinfo{person}{Huan Yan}, \bibinfo{person}{Can Rong}, \bibinfo{person}{Yuan Yuan}, \bibinfo{person}{Fenyu Jiang}, \bibinfo{person}{Zhenyu Han}, \bibinfo{person}{Hongjie Sui}, \bibinfo{person}{Depeng Jin}, {and} \bibinfo{person}{Yong Li}.} \bibinfo{year}{2024}\natexlab{b}.
\newblock \showarticletitle{Multi-scale simulation of complex systems: a perspective of integrating knowledge and data}.
\newblock \bibinfo{journal}{\emph{Comput. Surveys}} \bibinfo{volume}{56}, \bibinfo{number}{12} (\bibinfo{year}{2024}), \bibinfo{pages}{1--38}.
\newblock


\bibitem[Wang et~al\mbox{.}(2024a)]%
        {wang2024p}
\bibfield{author}{\bibinfo{person}{Qi Wang}, \bibinfo{person}{Pu Ren}, {et~al\mbox{.}}} \bibinfo{year}{2024}\natexlab{a}.
\newblock \showarticletitle{P2C2Net: PDE-Preserved Coarse Correction Network for efficient prediction of spatiotemporal dynamics}.
\newblock \bibinfo{journal}{\emph{arXiv preprint arXiv:2411.00040}} (\bibinfo{year}{2024}).
\newblock


\bibitem[Wang et~al\mbox{.}(2020)]%
        {wang2020deep}
\bibfield{author}{\bibinfo{person}{Yating Wang}, \bibinfo{person}{Siu~Wun Cheung}, \bibinfo{person}{Eric~T Chung}, \bibinfo{person}{Yalchin Efendiev}, {and} \bibinfo{person}{Min Wang}.} \bibinfo{year}{2020}\natexlab{}.
\newblock \showarticletitle{Deep multiscale model learning}.
\newblock \bibinfo{journal}{\emph{J. Comput. Phys.}}  \bibinfo{volume}{406} (\bibinfo{year}{2020}), \bibinfo{pages}{109071}.
\newblock


\bibitem[Waswani et~al\mbox{.}(2017)]%
        {waswani2017attention}
\bibfield{author}{\bibinfo{person}{A Waswani}, \bibinfo{person}{N Shazeer}, \bibinfo{person}{N Parmar}, \bibinfo{person}{J Uszkoreit}, \bibinfo{person}{L Jones}, \bibinfo{person}{A Gomez}, \bibinfo{person}{L Kaiser}, {and} \bibinfo{person}{I Polosukhin}.} \bibinfo{year}{2017}\natexlab{}.
\newblock \showarticletitle{Attention is all you need}. In \bibinfo{booktitle}{\emph{NIPS}}.
\newblock


\bibitem[Weinan(2011)]%
        {weinan2011principles}
\bibfield{author}{\bibinfo{person}{E Weinan}.} \bibinfo{year}{2011}\natexlab{}.
\newblock \bibinfo{booktitle}{\emph{Principles of multiscale modeling}}.
\newblock \bibinfo{publisher}{Cambridge University Press}.
\newblock


\bibitem[Woo et~al\mbox{.}(2018)]%
        {cbam_ref}
\bibfield{author}{\bibinfo{person}{Sanghyun Woo}, \bibinfo{person}{Jongchan Park}, \bibinfo{person}{Joon-Young Lee}, {and} \bibinfo{person}{In~So Kweon}.} \bibinfo{year}{2018}\natexlab{}.
\newblock \showarticletitle{CBAM: Convolutional Block Attention Module}.
\newblock \bibinfo{journal}{\emph{ECCV}} (\bibinfo{year}{2018}).
\newblock


\bibitem[Wu et~al\mbox{.}(2024b)]%
        {wu2024neural}
\bibfield{author}{\bibinfo{person}{Hao Wu}, \bibinfo{person}{Kangyu Weng}, \bibinfo{person}{Shuyi Zhou}, \bibinfo{person}{Xiaomeng Huang}, {and} \bibinfo{person}{Wei Xiong}.} \bibinfo{year}{2024}\natexlab{b}.
\newblock \showarticletitle{Neural Manifold Operators for Learning the Evolution of Physical Dynamics}. In \bibinfo{booktitle}{\emph{Proceedings of the 30th ACM SIGKDD Conference on Knowledge Discovery and Data Mining}}. \bibinfo{pages}{3356--3366}.
\newblock


\bibitem[Wu et~al\mbox{.}(2024a)]%
        {wu2024predicting}
\bibfield{author}{\bibinfo{person}{Tao Wu}, \bibinfo{person}{Xiangyun Gao}, \bibinfo{person}{Feng An}, \bibinfo{person}{Xiaotian Sun}, \bibinfo{person}{Haizhong An}, \bibinfo{person}{Zhen Su}, \bibinfo{person}{Shraddha Gupta}, \bibinfo{person}{Jianxi Gao}, {and} \bibinfo{person}{J{\"u}rgen Kurths}.} \bibinfo{year}{2024}\natexlab{a}.
\newblock \showarticletitle{Predicting multiple observations in complex systems through low-dimensional embeddings}.
\newblock \bibinfo{journal}{\emph{Nature Communications}} \bibinfo{volume}{15}, \bibinfo{number}{1} (\bibinfo{year}{2024}), \bibinfo{pages}{2242}.
\newblock


\bibitem[Xiao et~al\mbox{.}(2022)]%
        {xiao2022coupled}
\bibfield{author}{\bibinfo{person}{Xiongye Xiao}, \bibinfo{person}{Defu Cao}, \bibinfo{person}{Ruochen Yang}, \bibinfo{person}{Gaurav Gupta}, \bibinfo{person}{Gengshuo Liu}, \bibinfo{person}{Chenzhong Yin}, \bibinfo{person}{Radu Balan}, {and} \bibinfo{person}{Paul Bogdan}.} \bibinfo{year}{2022}\natexlab{}.
\newblock \showarticletitle{Coupled multiwavelet operator learning for coupled differential equations}. In \bibinfo{booktitle}{\emph{The Eleventh International Conference on Learning Representations}}.
\newblock


\bibitem[Yang et~al\mbox{.}(2023)]%
        {yang2023diffusion}
\bibfield{author}{\bibinfo{person}{Ling Yang}, \bibinfo{person}{Zhilong Zhang}, \bibinfo{person}{Yang Song}, \bibinfo{person}{Shenda Hong}, \bibinfo{person}{Runsheng Xu}, \bibinfo{person}{Yue Zhao}, \bibinfo{person}{Wentao Zhang}, \bibinfo{person}{Bin Cui}, {and} \bibinfo{person}{Ming-Hsuan Yang}.} \bibinfo{year}{2023}\natexlab{}.
\newblock \showarticletitle{Diffusion models: A comprehensive survey of methods and applications}.
\newblock \bibinfo{journal}{\emph{Comput. Surveys}} \bibinfo{volume}{56}, \bibinfo{number}{4} (\bibinfo{year}{2023}), \bibinfo{pages}{1--39}.
\newblock


\bibitem[Zang and Wang(2020)]%
        {zang2020neural}
\bibfield{author}{\bibinfo{person}{Chengxi Zang} {and} \bibinfo{person}{Fei Wang}.} \bibinfo{year}{2020}\natexlab{}.
\newblock \showarticletitle{Neural dynamics on complex networks}. In \bibinfo{booktitle}{\emph{Proceedings of the 26th ACM SIGKDD international conference on knowledge discovery \& data mining}}. \bibinfo{pages}{892--902}.
\newblock


\end{thebibliography}

\balance
\appendix
\setcounter{figure}{0}
\setcounter{table}{0}

\section{Experiments Setup}\label{app:exper_setup}

\subsection{Data Generation}
Here, we introduce the dynamics and data generation process for each complex system.
\textbf{Lambda–Omega system} is governed by 
\begin{equation}
\left\{
    \begin{aligned}
        \dot{u}_t&=\mu_u \Delta u + (1 - u^2 - v^2)u + \beta(u^2 + v^2)v \\
        \dot{v}_t&=\mu_v \Delta v + (1 - u^2 - v^2)v - \beta(u^2 + v^2)u,
    \end{aligned}
\right.
\end{equation}
where $\Delta$ is the  Laplacian operator.

\noindent \textbf{Brusselator system} is governed by 
\begin{equation}
\left\{
    \begin{aligned}
        \dot{u}_t&=\mu_u \Delta u + \alpha - (1 + \beta)u + u^2v \\
        \dot{v}_t&=\mu_v \Delta v + \beta u - u^2v,
    \end{aligned}
\right.
\end{equation}
while \textbf{Gray-Scott system} is governed by 
\begin{equation}
\left\{
    \begin{aligned}
        \dot{u}_t&=\mu_u \Delta u - uv^2 + \alpha(1 - u) \\
        \dot{v}_t&=\mu_v \Delta v + uv^2 - \beta v .
    \end{aligned}
\right.
\end{equation}

\noindent \textbf{Cylinder flow system} is governed by:
\begin{equation}
\left\{
    \begin{aligned}
        \dot{u}_t &= -u \cdot \nabla u - \frac{1}\alpha \nabla p + \frac\beta\alpha \Delta u, \\
        \dot{v}_t &= -v \cdot \nabla v + \frac{1}\alpha \nabla p - \frac\beta\alpha \Delta v .
    \end{aligned}
\right.
\end{equation}
The coefficient values and simulation settings for each equation are listed in Table~\ref{app_tab:simulation}.
All trajectories in the above systems are simulated from different initial conditions. The NS system data is sourced from \citet{takamoto2022pdebench}'s open repository, with 50 trajectories used for training and 12 for testing.
For the LO and Brusselator systems, the time is downsampled by a factor of 10, while for the GS system, the spatial resolution is interpolated to a $64 \times 64$ grid.

The cylinder flow system is simulated using the lattice Boltzmann method (LBM)~\cite{vlachas2022multiscale}, with dynamics governed by the Navier-Stokes equations for turbulent flow around a cylindrical obstacle. The system is discretized using a lattice velocity grid, and the relaxation time is determined based on the kinematic viscosity and Reynolds number. The spatial resolution is interpolated to a $128 \times 64$ grid, while the time is downsampled by a factor of 300. Data collection begins once the turbulence has stabilized.
We generate 50 training trajectories and 20 testing trajectories using varying Reynolds numbers, with 10 training trajectories having Reynolds numbers in the range [100, 500] and 10 out-of-distribution (OOD) trajectories in the range [500, 1000]. Using the formula \(\mu = \frac{\rho U_m D}{\text{Re}}\), where \(\rho = 1\), \(U_m = 0.08\), and \(D = 0.2\), the viscosities \(\mu\) for the training and OOD sets are calculated as \(\mu \in [3.2 \times 10^{-5}, 1.6 \times 10^{-4}]\) and \(\mu \in [1.6 \times 10^{-5}, 3.2 \times 10^{-5}]\), respectively.

Finally, we perform Min-max normalization along the channel dimension as the only data preprocessing.

\begin{table}[!h]
    \renewcommand{\arraystretch}{1.15}
    \centering
    \caption{Coefficient and settings of each system.}
    \resizebox{\linewidth}{!}{%
    \begin{tabular}{c|ccccccc}
        \hline \hline
         & $\mu_u$ & $\mu_v$ & $\alpha$ & $\beta$ & $dt$ & $T$ & spatial grids \\ \hline
        LO & 0.1 & 0.1 & — & 1.0 & 0.04 & 40.0 & $64 \times 64$ \\ 
        \hline
        Bruss & 1.0 & 0.1 & 1.0 & 3.0 & 0.02 & 20.0 & $64 \times 64$ \\ 
        \hline
        GS & $2 \times 10^{-5}$ & $1 \times 10^{-5}$ & 0.04 & 0.1 & 50.0 & $5 \times 10^{3}$ & $100 \times 100$ \\ 
        \hline
        CY & — & — & 1.0 & $\mu$ & 1.0 & $6 \times 10^4$ & $420 \times 180$ \\ 
        \hline \hline
    \end{tabular}%
    }
    \label{app_tab:simulation}
\end{table}

\subsection{Evaluation Metrics}

We use two metrics to evaluate the performance of our model: Normalized Mean Squared Error (NMSE) and Structural Similarity Index (SSIM).
The NMSE is computed as follows:
\begin{equation}
\text{NMSE} = \frac{\sum_{i=1}^{n} (y_i - \hat{y}_i)^2}{\sum_{i=1}^{n} y_i^2}
\end{equation}
where \( y_i \) represents the ground truth values, which has already been normalized and \( \hat{y}_i \) denotes the predicted values.

SSIM is a perceptual metric that measures the similarity between two signals. It is computed as:

\begin{equation}
\text{SSIM}(x, y) = \frac{(2 \mu_x \mu_y + c_1)(2 \sigma_{xy} + c_2)}{(\mu_x^2 + \mu_y^2 + c_1)(\sigma_x^2 + \sigma_y^2 + c_2)}
\end{equation}
where \( \mu_x, \mu_y \) are the means of \( x \) and \( y \), \( \sigma_x^2, \sigma_y^2 \) are the variances of \( x \) and \( y \), and \( \sigma_{xy} \) is the covariance between \( x \) and \( y \). The constants \( c_1 = 0.01^2 \) and \( c_2 = 0.03^2 \) are used to stabilize the division in the SSIM formula. For both NMSE and SSIM, the metrics are calculated for each snapshot, with the mean and standard deviation computed across the prediction time dimension and different trajectories to summarize the results.

\begin{figure}[!ht]
    \centering
    \includegraphics[width=0.9\linewidth]{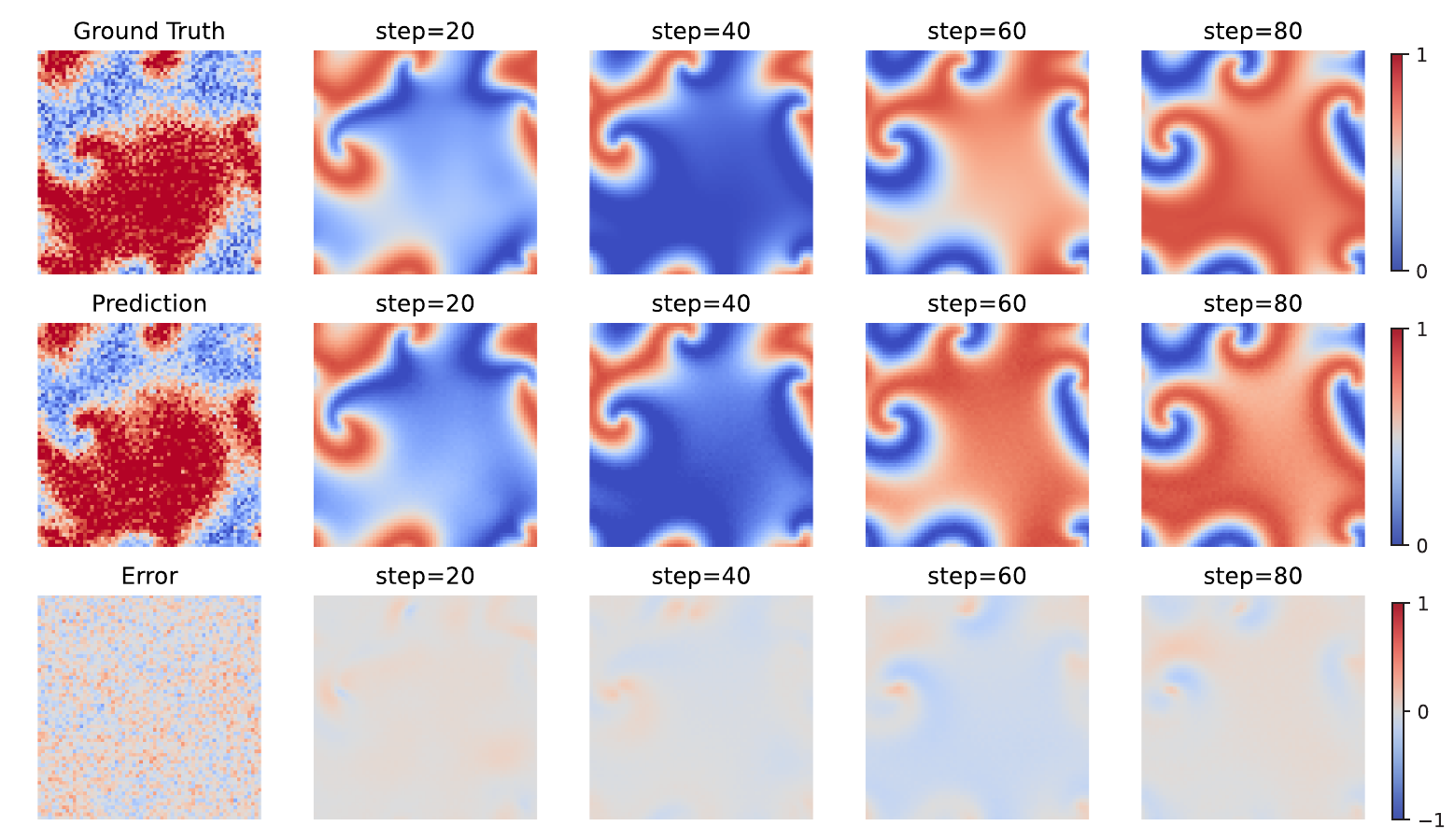}
    \caption{Snapshots of MDPNet's prediction results on LO system.}
    \label{apx:fig_lo}
\end{figure}
\begin{figure}[!ht]
    \centering
    \includegraphics[width=0.9\linewidth]{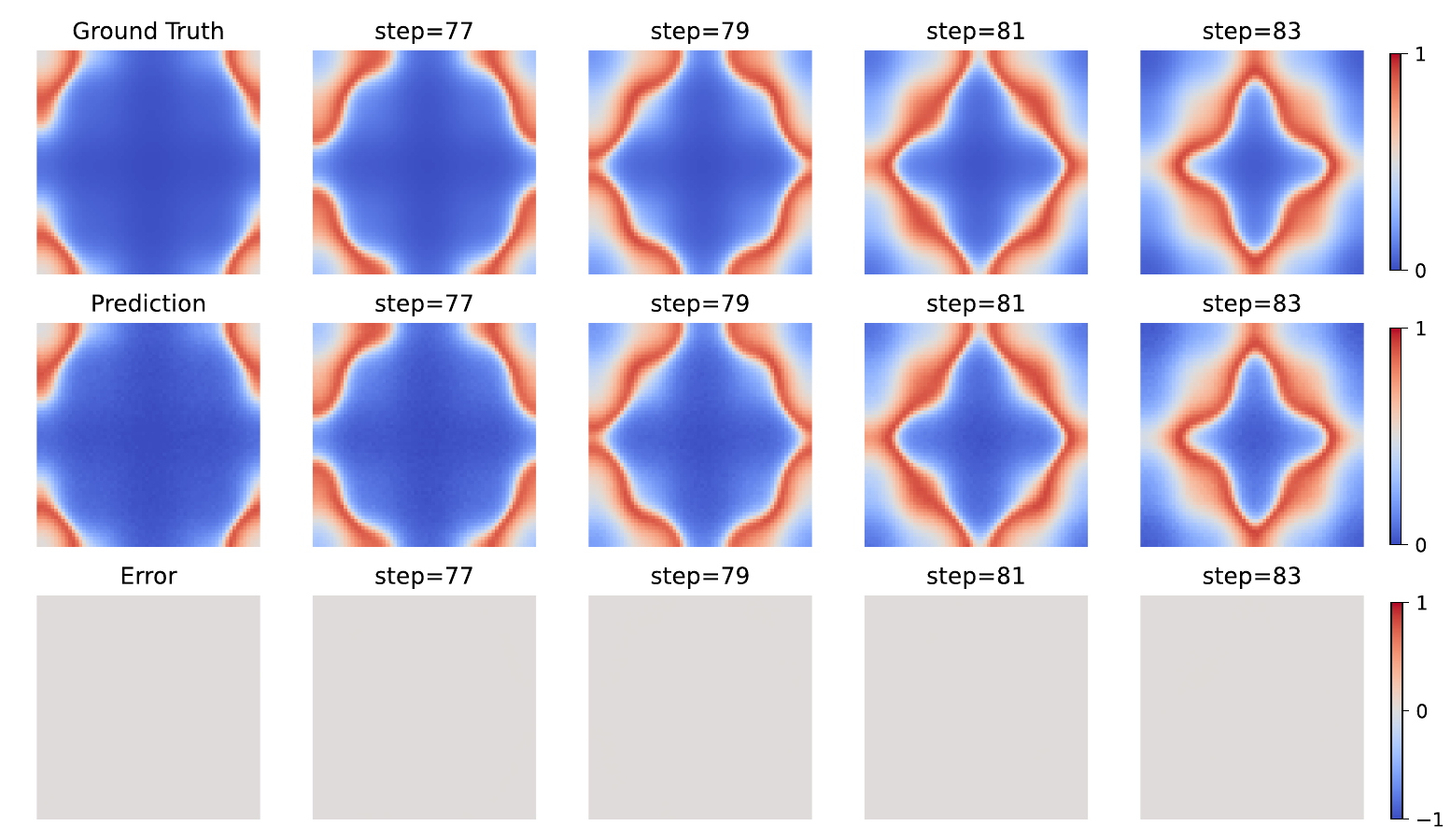}
    \caption{Snapshots of MDPNet's prediction results on Bruss system.}
    \label{apx:fig_bruss}
\end{figure}
\begin{figure}[!ht]
    \centering
    \includegraphics[width=0.9\linewidth]{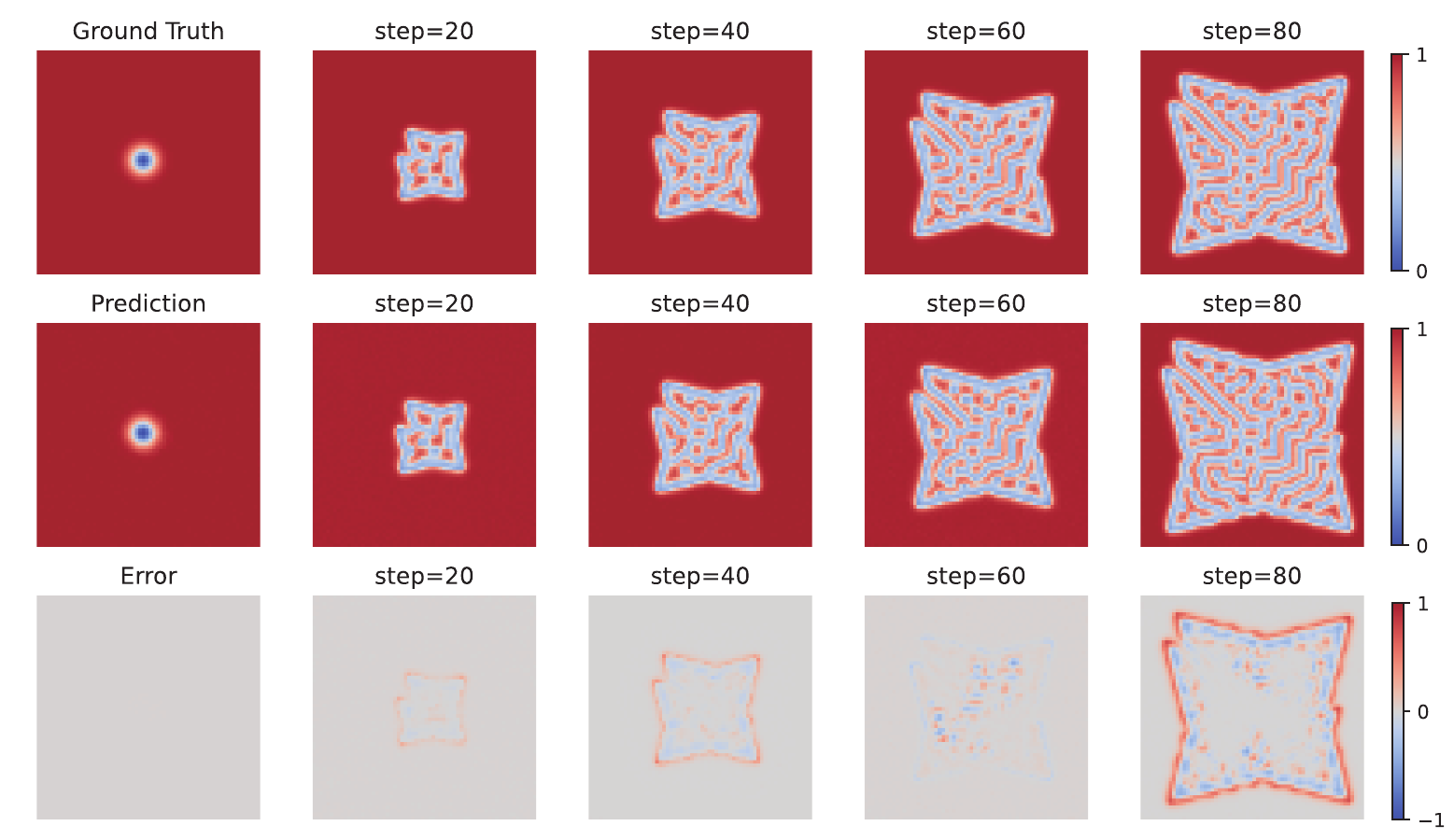}
    \caption{Snapshots of MDPNet's prediction results on GS system.}
    \label{apx:fig_gs}
\end{figure}
\begin{figure}[!ht]
    \centering
    \includegraphics[width=0.9\linewidth]{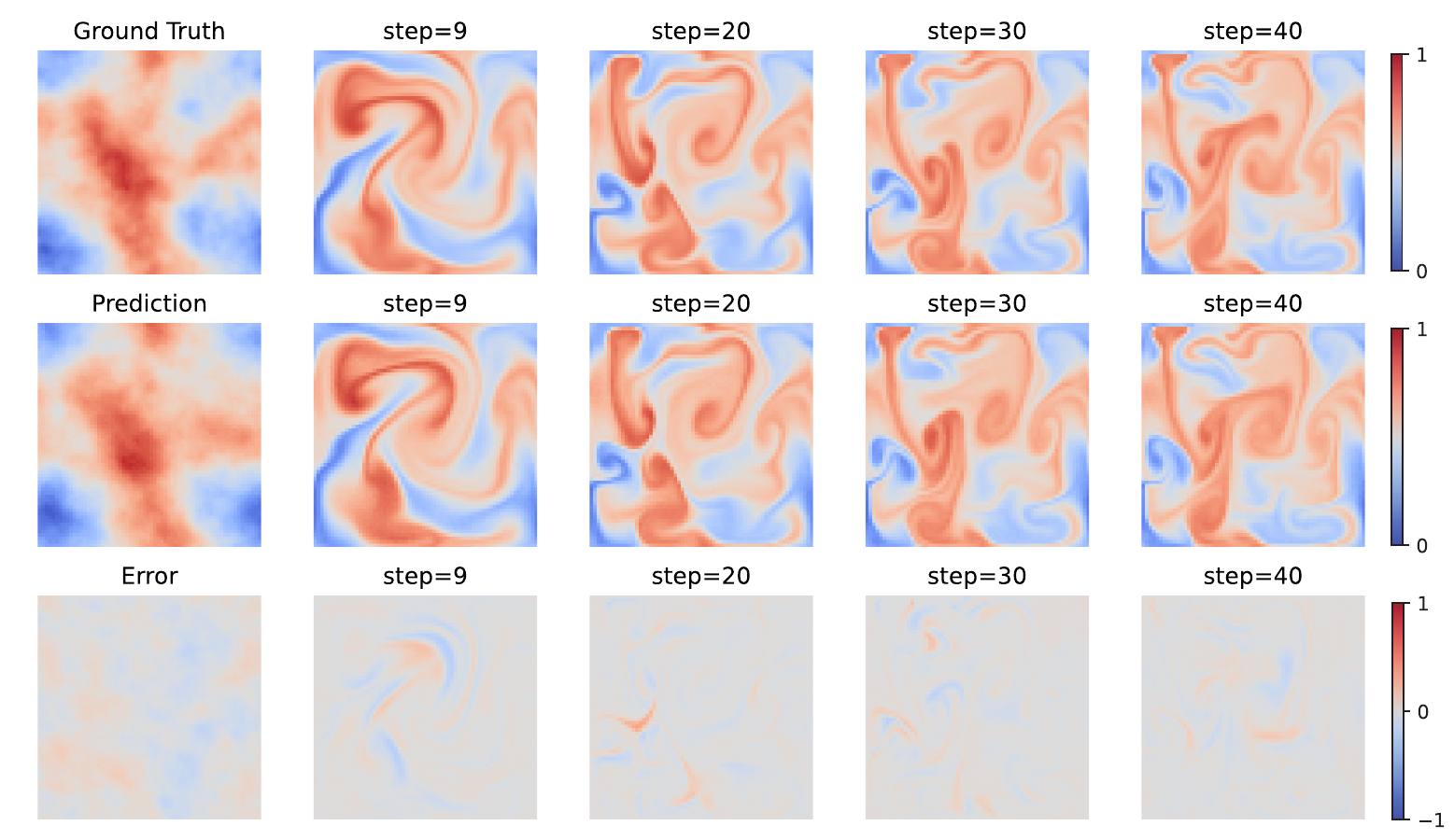}
    \caption{Snapshots of MDPNet's prediction results on NS system.}
    \label{apx:fig_ns}
\end{figure}

\section{Additional Results}\label{app:exper_add}

\subsection{Prediction Snapshots}
We visualize the comparison snapshots of long-term predicted trajectories and ground truth for MDPNet across four systems in Figures~\ref{apx:fig_lo}, \ref{apx:fig_bruss}, \ref{apx:fig_gs} and \ref{apx:fig_ns}. For the Bruss system, a segment of a limit cycle after long-term evolution is selected, while for the other systems, equal step-length sampling is used.

\subsection{Residual Snapshots}\label{Sec:apx_scale_residual}
\begin{figure}[!ht]
\centering
\subfigure[3-scale]{
\includegraphics[width=0.57\linewidth]{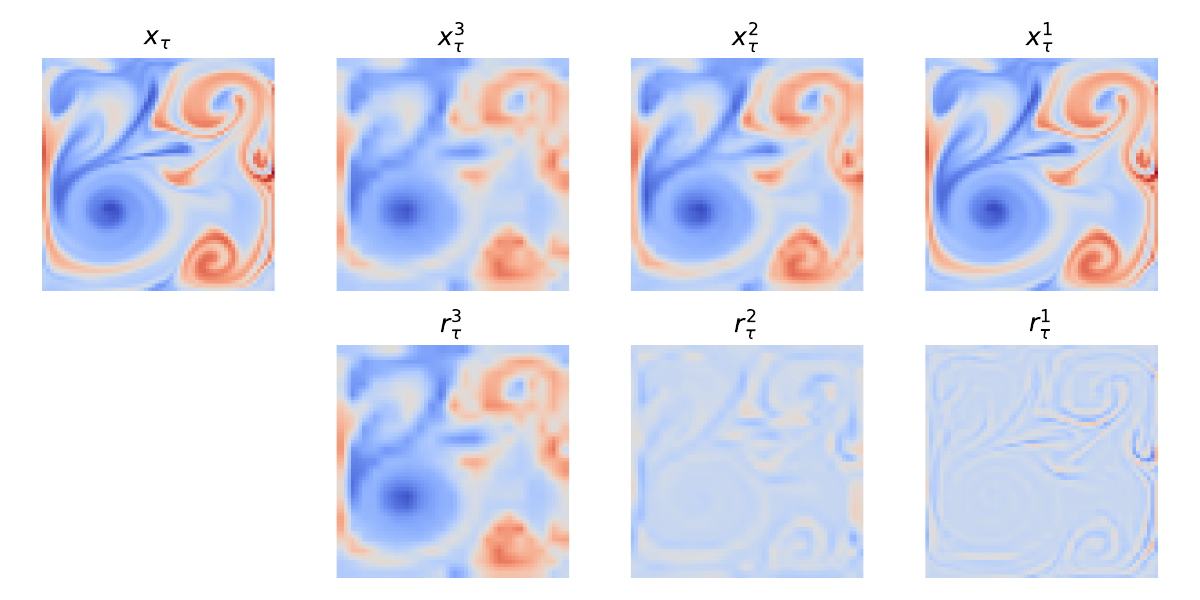}}
\subfigure[5-scale]{
\includegraphics[width=0.85\linewidth]{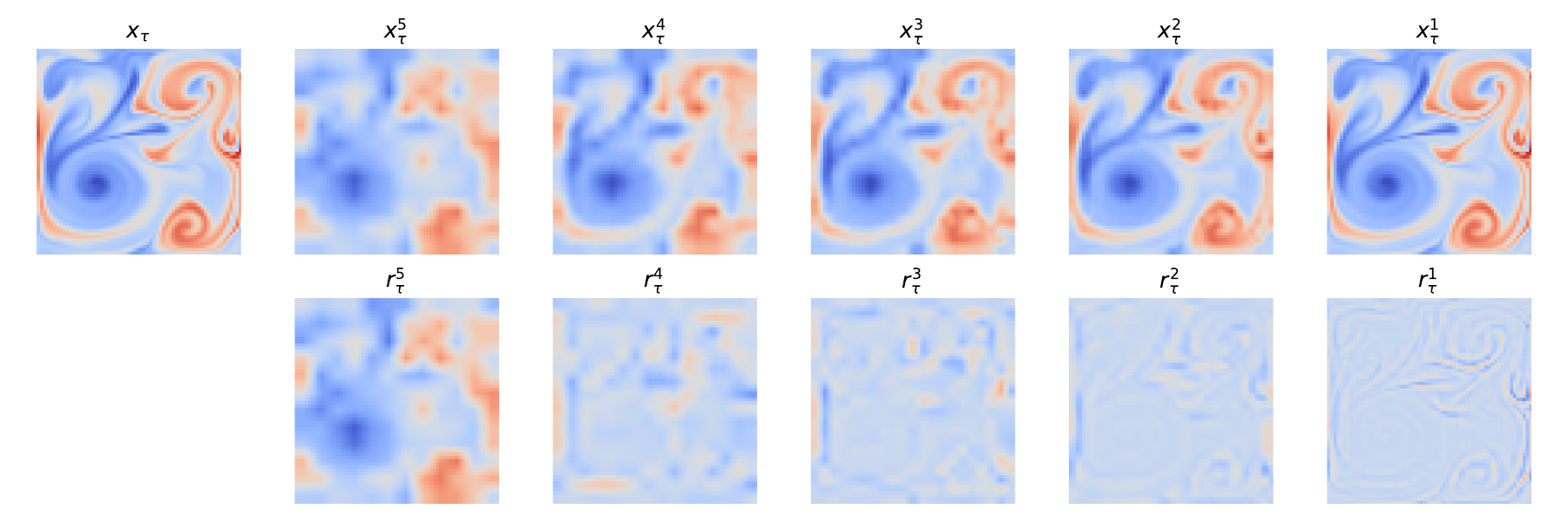}}
\caption{Snapshots of the residuals and coarse-graining at different scales in the NS system.}
\label{apx:residual}
\end{figure}

Using the NS system as an example, we show the residuals $r^k_\tau$ and coarse-grained states  $x^k_\tau$ at different scales, as shown in Figure~\ref{apx:residual}.
The residuals at each scale in the 3-scale model capture low, medium, and high-frequency components. The residual at the first scale in the 5-scale model is similar to that in the 3-scale model, but the amount of information related to the vortex distribution in the fifth scale residual is significantly reduced.

% \subsection{High-order System}

% To further evaluate MDPNet's modeling capabilities, we introduce a high-order (4th-order) system: the Swift–Hohenberg equation~\cite{swift1977hydrodynamic}. 
% This equation, given by 
% \begin{equation}
%     \frac{\partial u}{\partial t} = ru - (1+\Delta)^2 u + gu^2 - u^3,
% \end{equation}
% describes wrinkling morphologies and patterns in curved elastic bilayer materials. 
% We maintained the same data generation and preprocessing pipeline as before and compared the predictive performance of representative models. 
% As shown in Table~\ref{tab:sh_system}, MDPNet significantly outperforms other strong baselines, demonstrating its powerful modeling and predictive ability for high-order systems.

\begin{table}[!ht]
    \centering
    \caption{Prediction performance on the Swift–Hohenberg equation.}
    \begin{tabular}{lcc}
        \hline
        & \textbf{NMSE$\times 10^{-2}$} & \textbf{SSIM$\times 10^{-1}$} \\
        \hline
        ConvLSTM & 5.214 & 8.090 \\
        G-LED & 3.750 & 9.087 \\
        Ours & \textbf{2.043} & \textbf{9.477} \\
        \hline
    \end{tabular}
    \label{tab:sh_system}
\end{table}

\subsection{Ablation study on the diffusion step}

The original method uses uniform diffusion steps. 
We compare two heuristics: one based on the ratio of downsampled areas across scales, allocating steps from small to large (1:4:9) and from large to small (9:4:1). 
With 1,000 diffusion steps and 3 scales, uniform distribution (1:1:1) performs best (Table~\ref{tab:diffusion_step}), ensuring each scale gets sufficient attention.

% Requires: \usepackage{graphicx}
\begin{table}[h]
    \centering
    \caption{Different diffusion processes.}
    \begin{tabular}{cccccc}
        \hline
        NMSE$\times 10^{-2}$ & LO    & Bruss & GS    & NS    \\
        \hline
        1:4:9     & 2.444 & 0.179 & 2.841 & 1.454 \\
        9:4:1     & 2.682 & 0.244 & 3.227 & 1.241 \\
        1:1:1     & \textbf{1.145} & \textbf{0.044} & \textbf{1.144} & \textbf{1.154} \\
        \hline
    \end{tabular}
    \label{tab:diffusion_step}
\end{table}

\section{Model Architecture}

Our architecture combines established components from generative modeling and graph learning:

\subsection{Core Modules}
\begin{itemize}
    \item \textbf{Residual encoder $\phi_k$:} A deep CNN with residual blocks and a linear output. Each block includes Conv, GroupNorm, SiLU, AvgPool, followed by a CBAM \cite{cbam_ref}.
    \item \textbf{Diffusion decoder $\psi$:} A UNet with conditional attention. Each layer contains multiple residual attention blocks. In the upsampling path, each residual attention block is followed by a conditional attention block, where $q$ is computed from the latent vector $z_{\tau_k}$, and $k$, $v$ are computed from the feature vector $x_{\tau,n_k}$.
    \item \textbf{Predictor-self $f_{\theta_k}$:} A shared two-layer MLP with scale info in positional embeddings.
    \item \textbf{Predictor-inter $g_{\theta}$:} A standard graph attention layer \cite{gat_ref}.
\end{itemize}

\subsection{Model Hyperparameters}
\begin{itemize}
    \item \textbf{Depth:} $\phi_k$ has 2 residual blocks. $\psi$ has 3 layers, each with 2 residual attention blocks.
    \item \textbf{Width:} Feature dimension for $\phi_k$ and predictors is 128. Diffusion decoder channels: [64, 128, 128].
\end{itemize}

\subsection{Training Hyperparameters}
We train the model for 1000 epochs using a learning rate of 1e-4 and a batch size of 18. The ODE solver is \texttt{dopri5}.

\end{document}